\RequirePackage{fix-cm}
\documentclass[smallextended]{svjour3}       
\smartqed  
\usepackage{graphicx}
\newcommand{\bea}{\begin{eqnarray}}
\newcommand{\eea}{\end{eqnarray}}
\newcommand{\ben}{\begin{equation}}
\newcommand{\een}{\end{equation}}

\newcommand{\bes}{\begin{subequations}}
\newcommand{\ees}{\end{subequations}}
\usepackage{graphicx,amsmath,times,hhline}
\usepackage{amssymb,graphicx}
\usepackage{amsmath,bm}

\newcommand{\e}{\mbox{e}}
\newcommand{\al}{\alpha}
\newcommand{\ba}{\beta}
\newcommand{\del}{\delta}
\newcommand{\Del}{\Delta}
\newcommand{\ga}{\gamma}
\newcommand{\sech}{\mbox{ sech}}
 \journalname{NonlinearDyn}
\begin{document}

\title{Degenerate soliton solutions and their dynamics in the nonlocal Manakov system: I Symmetry preserving and symmetry breaking solutions
}


\author{S. Stalin    \and
        M. Senthilvelan   \and 
	M. Lakshmanan
}


\institute{S. Stalin \at
             Centre for Nonlinear Dynamics, 
	      School of Physics, 
	      Bharathidasan University, 
		Tiruchirappalli-620 024, 
		Tamil Nadu, India
           \and
           M. Senthilvelan \at
              Centre for Nonlinear Dynamics, 
	      School of Physics, 
	      Bharathidasan University, 
Tiruchirappalli-620 024, 
		Tamil Nadu, India\\
              \email{velan@cnld.bdu.ac.in}
	\and
           M. Lakshmanan \at
              Centre for Nonlinear Dynamics, 
	      School of Physics, 
	      Bharathidasan University, 
		Tiruchirappalli-620 024, 
		Tamil Nadu, India\\
              \email{lakshman@cnld.bdu.ac.in} 
}

\date{Received: date / Accepted: date}

\maketitle

\begin{abstract}

In this paper, we construct degenerate soliton solutions (which preserve $\cal{PT}$-symmetry/break $\cal{PT}$-symmetry) to the nonlocal Manakov system through a nonstandard bilinear procedure. 
 Here by degenerate we mean the solitons that are present in both the modes which propagate with same velocity. The degenerate nonlocal soliton solution is constructed after briefly indicating the form of nondegenerate one-soliton solution. To derive these soliton solutions, we simultaneously solve the nonlocal Manakov equation and a pair of coupled equations that arise from the zero curvature condition. The later consideration yields general soliton solution which agrees with the solutions that are already reported in the literature under certain specific parametric choice. We also discuss the salient features associated with the obtained degenerate soliton solutions.
\end{abstract}

\section{Introduction}
\label{intro}
In the context of $\cal{PT}$-symmetric classical optics \cite{a,b}, recently a nonlocal nonlinear Schr\"{o}dinger (NNLS) equation, namely 
\bea
iq_{t}(x,t)+q_{xx}(x,t)+2\sigma q(x,t)q^{*}(-x,t)q(x,t)=0, ~\sigma=\pm 1.
\label{a}
\eea
has been introduced in \cite{3a}. It has been shown that Eq. (\ref{a}) is completely integrable \cite{3a,3b,4a}, since it admits a Lax pair, infinite number of conservation laws and is solvable by inverse scattering transform (IST) technique. Eq. (\ref{a}) has a self induced-potential $V(x,t)=2\sigma q(x,t)q^{*}(-x,t)$ which obeys the $\cal{PT}$-symmetry condition $V^*(-x,t)= V(x,t)$ \cite{1}. In Eq. (\ref{a}), if we replace the nonlocal $q^{*}(-x,t)$ term by $q^{*}(x,t)$, it becomes standard NLS equation. The nonlocal term in (\ref{a}) implies that the field at $x$ always requires information from the field at $-x$ \cite{3} simultaneously. That is the field $q^{*}(-x,t)$ is either independent or dependent with respect to the field $q(x,t)$. In the dependent case, the field  $q^{*}(-x,t)$ is a parity conjugate of $q(x,t)$ in which the solution exhibits $\cal{PT}$-symmetry while in the independent case the function $q^{*}(-x,t)$ is not parity transformed ($x\rightarrow -x$) complex conjugate function of  $q(x,t)$ which corresponds to $\cal{PT}$-symmetry broken case. The NNLS equation is gauge equivalent to an unconventional system of coupled Landau-Lifshitz equation \cite{4b}. In contradiction to this the standard NLS equation has been shown to be $L$-equivalent to Heisenberg spin chain equation in the continuum limit \cite{4c}. Various recent studies have shown that the analysis of NNLS equation and its variants have become one of the active areas of research both from physical and mathematical perspectives \cite{5a}-\cite{67}. 

The non-trivial generalization of Eq. (\ref{a}) is the vector nonlocal NLS equation or coupled NNLS equation, namely
\bea
iq_{j,t}(x,t)+q_{j,xx}(x,t)&+&2\sum_{l=1}^{2}\sigma_{l}q_l(x,t)q_l^{*}(-x,t)q_{j}(x,t)=0, ~j=1,2.
\label{1.1a}
\eea
In Eq. (\ref{1.1a}), each $q_j(x,t)$ is a complex valued wave envelope and $q_{j,t}$ and $q_{j,x}$ represent the derivatives of $q_j$ with respect to $t$ and $x$, respectively. In the above equation, $q^{*}_l(-x,t)$ are the nonlocal fields, and the local CNLS equation can be obtained by replacing it by local fields $q_l^{*}(x,t)$. In Eq. (\ref{1.1a}), the nonlocal version of self phase modulation and cross phase modulation constitute the nonlocal nonlinearity. Analogues to the local CNLS equations, Eq. (\ref{1.1a}) comprises of three different equations, namely focusing, defocusing and mixed type depending on the signs of the nonlinearity coefficients $\sigma_{l}$'s. If $\sigma_l=+1$, $l=1,2$, Eq. (\ref{1.1a}) becomes the focusing CNNLS equation or the nonlocal version of the celebrated local Manakov equation. The local Manakov equation is shown to possesses several interesting properties \cite{6a,6b}, including shape changing property of solitons under collision. When $\sigma_l=-1$, $l=1,2$, Eq. (\ref{1.1a}) becomes the defocusing coupled NNLS equation. Its local counterpart is the defocusing coupled NLS equation which admits dark-dark and bright-dark soliton solutions \cite{6c,6d}. Defocusing coupled NLS equation does not admit any shape changing property \cite{6d}. If $\sigma_l=\pm 1$ ($\sigma_1=+1$, $\sigma_2=-1$ and vice-versa), Eq. (\ref{1.1a}) becomes the coupled NNLS equation with mixed focusing-defocusing nonlinearity. The local version of it admits brigh-bright, bright-dark and dark-dark type soliton solutions \cite{7a}-\cite{7b}.  The above facts emphasize that to study the collision between solitons in the underlying nonlocal system, it is essential to derive multi-soliton solutions. 

In Ref, \cite{8a} the authors have obtained a two parameter family of breathing finite time blowup one soliton solution for the Eq. (\ref{1.1a}) with $\sigma_l=+1$. Very recently, soliton solutions have been constructed for the various coupled nonlocal field models by  applying non-vanishing boundary conditions. However, to the best of our knowledge, for the first time, we report in this paper bright one and two soliton solutions of the nonlocal Manakov equation, that is Eq. (\ref{1.1a}) with $\sigma_l=+1$, $l=1,2$. For convenience, we divide our investigation into two parts. In the present first part, we focus our attention only on the derivation of soliton solutions to the nonlocal Manakov equation. In the second subsequent part, we investigate the collision dynamics between the degenerate two- solitons in detail by using the obtained two soliton solution. 

 To explore general soliton solutions, we adopt the non-standard bilinearization procedure developed for the scalar NNLS equation \cite{5e}. Using this procedure, we  bilinearize both the nonlocal Manakov equation and the following a pair of coupled equations that arise in the zero curvature condition \cite{8a}, that is
\bea
iq_{j,t}^*(-x,t)-q_{j,xx}^*(-x,t)&-&2\sum_{l=1}^{2}q_l^{*}(-x,t)q_l(x,t)q_{j}^{*}(-x,t)=0, ~j=1,2.
\label{1.1b}
\eea
The reason behind the inclusion of the above equations in the solution construction process is that to introduce more number of complex parameters in the soliton solutions since the number of distinct eigenvalues arise in pair in one and higher order soliton solutions and the possibility of locating eigenvalues anywhere in the complex plane leads to new eigenvalue configuration in the nonlocal family of equations while solving the left/right Riemann-Hilbert problem. Due to the above reasons we treat the functions $q_j(x,t)$ and $q_j^*(-x,t)$, $j=1,2$ as independent entities. As we pointed out earlier, in the general case, the functions $q_j^*(-x,t)$ need not always the parity transformed complex conjugate of $q_j(x,t)$.

To bilinearize Eqs. (\ref{1.1a}) and (\ref{1.1b}), we introduce two auxiliary functions in the bilinear process in order to obtain the bilinear forms of them. By solving the obtained bilinear equations systematically, we derive degenerate one and two bright soliton solutions for the nonlocal Manakov equation. From the obtained one soliton solution, we match the solutions that already exist in the literature under certain parametric choice.   Besides deriving the one and two soliton solutions, we also discuss the salient features of the obtained soliton solutions. 

 The outline of the paper is as follows. In section 2, we describe the bilinearization of Eqs. (\ref{1.1a}) and  (\ref{1.1b}) using the nonstandard bilinear procedure. In Sec. 3, to begin with, we construct non-degenerate one soliton solution from which we extract the degenerate one soliton solution under specific restriction on the wavenumbers and discuss the salient features associated with it. In Sec. 4,  we derive the degenerate two soliton solutions of Eq. (\ref{1.1a}).   We also show that the obtained two-soliton solution can be reduced to a simple form.  We present our conclusions in Sec. 5.  
\section{Nonstandard bilinearization procedure}
\label{sec:1}
The nonlocal Manakov equation (\ref{1.1a}) is integrable and solvable by IST method \cite{8a}. In Ref. \cite{8a}, the authors have derived a two parameter family of breathing one soliton solution for Eq. (\ref{1.1a}) with $\sigma_l=+1$, $l=1,2$ through IST. However, to the best of our knowledge, explicit form of two soliton solution or higher order soliton solutions for this equation has not been reported so far. To capture the known solutions we have to modify the procedure appropriately. Interestingly, the  modified procedure generates  more general solutions for this equation. 

 As we pointed out earlier, in the bilinear process, we also incorporate Eq. (\ref{1.1b}) along with the Eq. (\ref{1.1a}). This augmentation is very much necessary to construct general soliton solutions. To bilinearize Eqs. (\ref{1.1a}) and (\ref{1.1b}) (with $\sigma_l=+1$, $l=1,2$) simultaneously we consider the following transformations, namely
\bea
q_j(x,t)=\frac{g^{(j)}(x,t)}{f(x,t)}, ~q_j^{*}(-x,t)=\frac{g^{(j)*}(-x,t)}{f^{*}(-x,t)}, ~j=1,2\label{2.1},
\eea     
where $g^{(j)}(x,t)$, $g^{(j)*}(-x,t)$, $f(x,t)$ and $f^{*}(-x,t)$ are all complex functions  and they are all considered as distinct to start with. To obtain the bilinear forms of (\ref{1.1a}) and (\ref{1.1b}) we introduce two auxiliary functions, one each for the coupled NNLS Eq. (\ref{1.1a}) and (\ref{1.1b}), respectively. By introducing equal number of auxiliary functions we can match the number of bilinear equations with equal number of unknown functions \cite{6,7} which in turn provides a  nontrivial consistent solution  to the given problem, as we see below.  
 
 Substituting the transformation given in (\ref{2.1}) in Eqs. (\ref{1.1a}) and (\ref{1.1b}), we obtain the following bilinear equations, that is
\bes\bea
D_1g^{(j)}(x,t)\cdot f(x,t)&=&2g^{(j)}(x,t) \cdot s^{(1)}(-x,t),\label{2.2a} \\
D_2f(x,t)\cdot f(x,t)&=&4s^{(1)}(-x,t) \cdot f(x,t), \label{2.2b}\\
D_3g^{(j)*}(-x,t)\cdot f^{*}(-x,t)&=&-2g^{(j)*}(-x,t) \cdot s^{(2)}(-x,t), \label{2.2c} \\
D_2f^{*}(-x,t)\cdot f^{*}(-x,t)&=&4s^{(2)}(-x,t) \cdot f^{*}(-x,t),~j=1,2,
\label{2.2d}
\eea\ees 
where $D_1\equiv(iD_t+D_x^2)$, $D_2\equiv D_x^2$, $D_3\equiv(iD_t-D_x^2)$ and $D_{t}$ and $D_{x}$ are the standard Hirota's bilinear operators \cite{8}. The auxiliary functions are defined by
\bes\bea
s^{(1)}(-x,t)\cdot f^{*}(-x,t)&=&\sum_{n=1}^{2}g^{(n)}(x,t)\cdot g^{(n)*}(-x,t),\label{2.3a}\\
s^{(2)}(-x,t)\cdot f(x,t)&=&\sum_{n=1}^{2}g^{(n)}(x,t)\cdot g^{(n)*}(-x,t).
\label{2.3b}
\eea\ees
The above set of bilinear Eqs. (5) can be solved by expanding the unknown functions $g^{(j)}(x,t)$, $g^{(j)*}(-x,t)$, $f(x,t)$, $f^{*}(-x,t)$, $s^{(1)}(-x,t)$ and $s^{(2)}(-x,t)$ in the following manner:
\bes\bea
&&g^{(j)}=\epsilon g_{1}^{(j)}+\epsilon^{3} g_{3}^{(j)}+..., ~g^{(j)*}=\epsilon g_{1}^{(j)*}+\epsilon^{3} g_{3}^{(j)*}+...,\label{2.4a}\\
&&f=1+\epsilon^{2} f_{2}+\epsilon^{4} f_{4}+...,~f^{*}=1+\epsilon^{2}f^{*}_{2}+\epsilon^{4}f^{*}_{4}+...,\label{2.4b}\\
&&s^{(1)}=\epsilon^{2}s^{(1)}_{2}+\epsilon^{4}s^{(1)}_{4}+...,~s^{(2)}=\epsilon^{2}s^{(2)}_{2}+\epsilon^{4}s^{(2)}_{4}+..,~j=1,2. \label{2.4c}
\eea\ees
Here, $\epsilon$ is a small expansion parameter. We can obtain a set of linear partial differential equations (PDEs) by collecting the coefficients of same powers of $\epsilon$ after substituting the above expansions in (\ref{2.2a})-(\ref{2.2d}). By solving them recursively we can obtain the explicit forms of the unknown functions appearing in (7). Substituting the relevant expressions back in (\ref{2.1}) we can get the soliton solutions of Eq. (\ref{1.1a}). We note here that in the conventional bilinearization procedure, for the local coupled NLS equation, we have only a pair of bilinear equations for the unknown functions $g^{(j)}(x,t)$, $j=1,2$, and $f(x,t)$ \cite{6a}. Here we have to find a consistent solution that satisfies all the six equations given in (5). 
\section{One-soliton solution}
To begin, we demonstrate the method of constructing nondegenerate and degenerate one-soliton solutions for Eqs. (\ref{1.1a}) and (\ref{1.1b}).
\subsection{Nondegenerate and Degenerate nonlocal one-soliton solution}
The solitons in which both the modes propagate with the same velocity are called degenerate solitons \cite{7f,7g}. To explore degenerate solitons in Eq. (\ref{1.1a}),  we begin our analysis with the following lowest order linear PDEs, that is
\bea
ig_{1t}^{(j)}(x,t)+g_{1xx}^{(j)}(x,t)=0, ~ig_{1t}^{(j)*}(-x,t)-g_{1xx}^{(j)*}(-x,t)=0, ~j=1,2.\label{lo}
\eea
The above Eqs. (\ref{lo}) admit the following solutions, namely
\bes
\bea
&&g^{(j)}_1(x,t)=\al^{(j)}_1\e^{\bar{\xi}_1^{(j)}},~\bar{\xi}_1^{(j)}=i \bar{k}_{1}^{(j)}x-i\bar{k}_{1}^{(j)^2}t,\label{nd1}\\
&&g^{(j)*}_1(-x,t)=\ba^{(j)}_1 \e^{\xi_1^{(j)}},~\xi_1^{(j)}=i k_{1}^{(j)}x+ik_{1}^{(j)^2}t, ~j=1,2. \label{nd2}
\eea \ees
In the above solutions one may notice that the exponential functions which are present in both the modes are different, that is the exponential functions in $g^{(1)}_1(x,t)$ and $g^{(2)}_1(x,t)$ are different. Similarly the exponential functions in the fields $g^{(1)*}_1(-x,t)$ and $g^{(2)*}_1(-x,t)$ are also different. This consideration leads to the solitons which propagate with different velocities in different modes. Such type of solitons are non-degenerate solitons. For example, proceeding with the forms given in Eqs. (\ref{nd1}) and (\ref{nd2}), we find that the series expansion (\ref{2.4a})-(\ref{2.4c}) get truncated for non-degenerate one soliton solution at $7$-th order in $g^{(j)}(x,t)$ and $g^{(j)*}(-x,t)$, at $8$-th order in $f(x,t)$ and $f^{*}(-x,t)$ and $6$-th order in $s^{(1)}(-x,t)$ and $s^{(2)}(-x,t)$.  Using these forms and substituting them in (\ref{2.1}), we obtain the expressions for one-soliton solution explicitly. 

 The factorized compact form of non-degenerate one-soliton solution can then be expressed as,  
\bes\bea
q_j(x,t)&=&\frac{\al_1^{(j)}\e^{\bar{\xi}_1^{(j)}}+\e^{\bar{\xi}_1^{(1)}+\bar{\xi}_1^{(2)}+\xi_1^{(3-j)}+\Del_1^{(j)}}}{1+\e^{\xi_1^{(1)}+\bar{\xi}_1^{(1)}+\del_1}+\e^{\xi_1^{(2)}+\bar{\xi}_1^{(2)}+\del_2}+\e^{\xi_1^{(1)}+\bar{\xi}_1^{(1)}+\xi_1^{(2)}+\bar{\xi}_1^{(2)}+\del_3}},\label{nod1}\\
q_j^*(-x,t)&=&\frac{\ba_1^{(j)}\e^{\xi_1^{(j)}}+\e^{\xi_1^{(1)}+\xi_1^{(2)}+\bar{\xi}_1^{(3-j)}+\ga_1^{(j)}}}{1+\e^{\xi_1^{(1)}+\bar{\xi}_1^{(1)}+\del_1}+\e^{\xi_1^{(2)}+\bar{\xi}_1^{(2)}+\del_2}+\e^{\xi_1^{(1)}+\bar{\xi}_1^{(1)}+\xi_1^{(2)}+\bar{\xi}_1^{(2)}+\del_3}},\label{nod2}
\eea \ees
where the explicit forms of the constants appearing in the above soliton solution are given in Appendix A.

  We point out that for constructing non-degenerate one-soliton solution itself requires analysis upto order of $\epsilon^8$. At this stage to proceed with the analysis of multi-soliton solutions for the non-degenerate case is too cumbersome. Therefore in this paper, we restrict ourselves to investigate the degenerate soliton solution only which is obtained from the above non-degenerate one-soliton solution. However, we plan to analyze the above non-degenerate one soliton solution in more detail and construct the corresponding multi-soliton solutions and study their dynamics in-detail in a follow-up work. In this and subsequent papers, we restrict ourselves to the construction of  degenerate soliton solutions for Eq. (\ref{1.1a}). Hence we impose a constraint on the wave numbers in the exponential functions in both the modes, that is the wave numbers are chosen to be $\bar{k}_{1}^{(1)}=\bar{k}_{1}^{(2)}=\bar{k}_{1}$ and $ k_{1}^{(1)}=k_{1}^{(2)}=k_{1}$. This restriction enforces the exponential functions in  $g^{(1)}_1(x,t)$ and $g^{(2)}_1(x,t)$ to be one and the same. Similarly the exponential functions in $g^{(1)*}_1(-x,t)$ and $g^{(2)*}_1(-x,t)$ are same. This restriction allows us to explore degenerate solitons in Eq. (\ref{1.1a}). As we demonstrate below even this degenerate soliton solutions reveal very interesting properties. 

Imposing the above said restriction on the wave numbers, we have the following expressions for the functions $g_1^{(j)}$ and $g_1^{(j)*}$, that is
\bes
\bea
&&g^{(j)}_1(x,t)=\al^{(j)}_1\e^{\bar{\xi}_1},~~\bar{\xi}_1=i \bar{k_{1}}x-i\bar{k_{1}^{2}}t, \label{deg1}\\
&&g^{(j)*}_1(-x,t)=\ba^{(j)}_1 \e^{\xi_1},~~\xi_1=i k_{1}x+ik_{1}^{2}t, ~j=1,2.\label{deg2} 
\eea\ees

 Now the modes differ from each other only in their (complex) amplitudes. The above restriction on the wave numbers enforces us to truncate the series expansion (\ref{2.4a})-(\ref{2.4c}) at $3$-rd order in $g^{(j)}(x,t)$ and $g^{(j)*}(-x,t)$, at $4$-th order in $f(x,t)$ and $f^{*}(-x,t)$ and $4$-th order in $s^{(1)}(-x,t)$ and $s^{(2)}(-x,t)$.  Consequently solving the system of resultant linear partial differential equations, which result from the bilinear equations, using the inputs (\ref{deg1})-(\ref{deg2}), we find  
\bes
\bea
&&g^{(j)}_3(x,t)=\e^{\xi_1+2\bar{\xi_1}+\Del_1^{(j)}},~g_3^{(j)*}(-x,t)=\e^{2\xi_1+\bar{\xi_1}+\ga_1^{(j)}},~\e^{\Del_1^{(j)}}=-\frac{\al^{(j)}_1\Gamma_{11}}{\kappa_{11}},~~~~\label{2.11a}\\
&&f_2(x,t)=f_2^{*}(-x,t)=\e^{\xi_1+\bar{\xi}_1+\del_1},~\e^{\del_1}=-2\frac{\Gamma_{11}}{\kappa_{11}},~ \e^{\ga_1^{(j)}}=-\frac{\ba^{(j)}_1\Gamma_{11}}{\kappa_{11}},~\label{2.9c}\\
&&f_4(x,t)=f^{*}_4(-x,t)=\e^{2(\xi_1+\bar{\xi_1})+R},~\e^{R}=\frac{\Gamma_{11}^2}{\kappa_{11}^2}, ~j=1,2,~\label{2.13}
\eea \ees
whereas  the auxiliary functions are reduced to
\bea
s^{(1)}_2(-x,t)=s^{(2)}_2(-x,t)=\Gamma_{11}\e^{\xi_1+\bar{\xi}_1}.
\eea
In the above $\kappa_{11}=(k_1+\bar{k}_1)^2$ and $\Gamma_{11}=(\al^{(1)}_1\ba^{(1)}_1+\al^{(2)}_1\ba^{(2)}_1)$. One can check that the auxiliary functions $s^{(1)}_4(-x,t)$ and $s^{(2)}_4(-x,t)$ become zero at the order of $\epsilon^4$. 

 Substituting the expressions found above in (\ref{2.1}), we arrive at the following degenerate one bright soliton solution, namely
\bes
\ben
q_j(x,t)=\frac{\al^{(j)}_1\e^{\bar{\xi}_1}+\e^{\xi_1+2\bar{\xi_1}+\Del_{j1}}}{1+\e^{\xi_1+\bar{\xi_1}+\del_1}+\e^{2(\xi_1+\bar{\xi_1})+R}}\equiv \frac{\al^{(j)}_1\e^{\bar{\xi}_1}}{1+\e^{\xi_1+\bar{\xi_1}+\Del}},~\e^{\Del}=-\frac{\Gamma_{11}}{\kappa_{11}}.
\label{2.14a}
\een
The fields $q^{*}_j(-x,t)$ turns out to be 
\ben
q^{*}_j(-x,t)=\frac{\ba^{(j)}_1\e^{\xi_1}+\e^{2\xi_1+\bar{\xi_1}+\ga_{j1}}}{1+\e^{\xi_1+\bar{\xi_1}+\del_1}+\e^{2(\xi_1+\bar{\xi_1})+R}}\equiv\frac{\ba^{(j)}_1\e^{\xi_1}}{1+\e^{\xi_1+\bar{\xi_1}+\Del}}.\label{2.14b}
\een
\ees

 It is a straightforward matter to verify the correctness of the solutions (\ref{2.14a}) and (\ref{2.14b}) by substituting them back in Eqs. (\ref{1.1a}) and (\ref{1.1b}). The one bright soliton solution given above is characterized by six complex parameters, namely $\al_1^{(j)}$, $\ba_1^{(j)}$, $j=1,2$, $k_1$ and $\bar{k}_1$, whereas the degenerate one bright soliton solution of local Manakov equation is characterized by only three complex parameters \cite{6a,6b,7a}. We note that the functions $q^{*}_j(-x,t)$ given in (\ref{2.14b}) are in general not parity conjugate of $q_j(x,t)$ given in  (\ref{2.14a}).   
\begin{figure}[ht]
\centering
\includegraphics[width=0.3\linewidth]{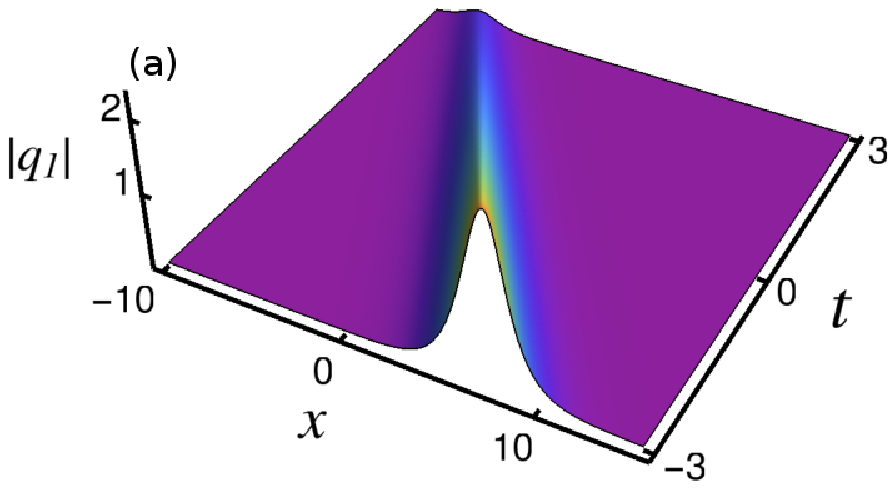}~~~\includegraphics[width=0.3\linewidth]{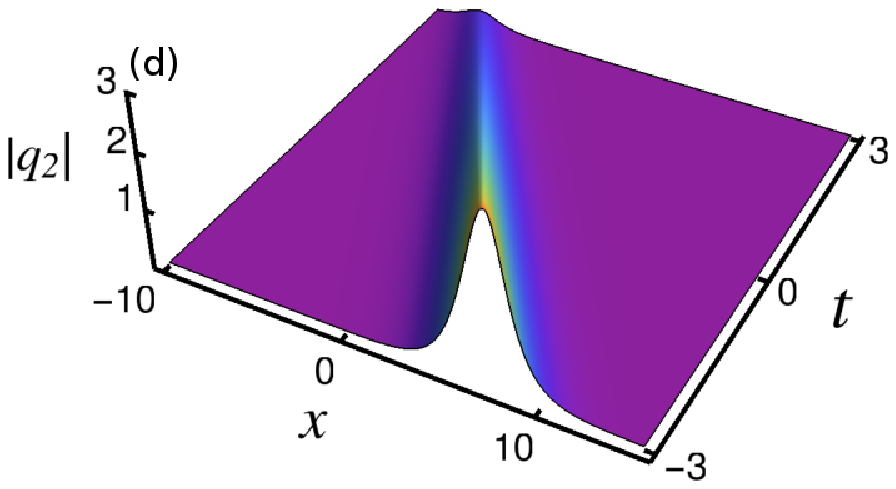}\\
\includegraphics[width=0.3\linewidth]{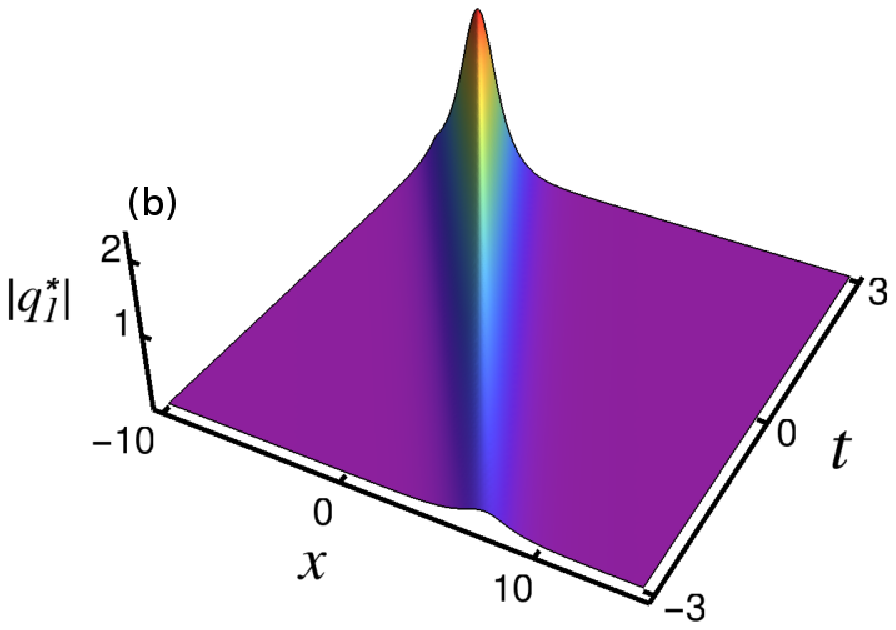}~~~\includegraphics[width=0.3\linewidth]{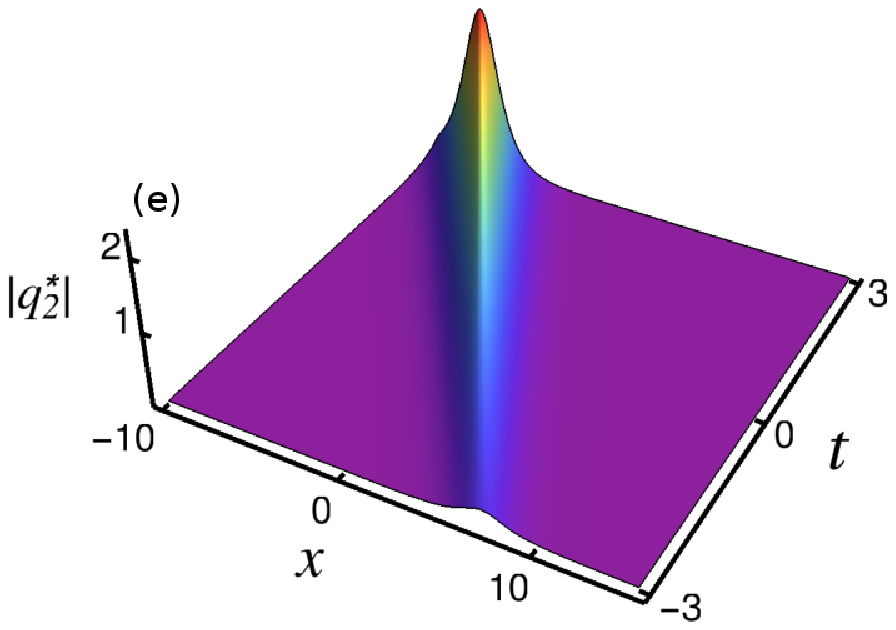}\\
\includegraphics[width=0.3\linewidth]{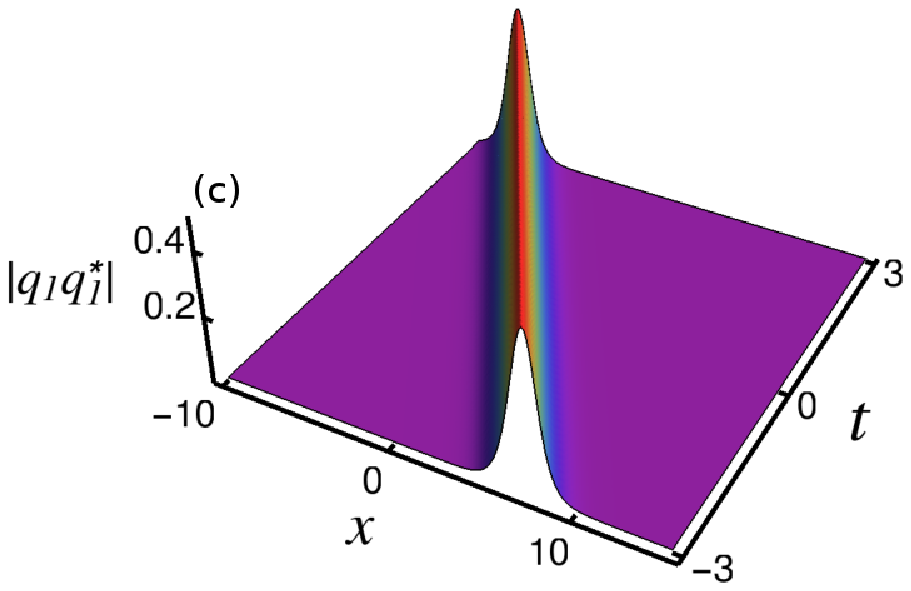}~~\includegraphics[width=0.3\linewidth]{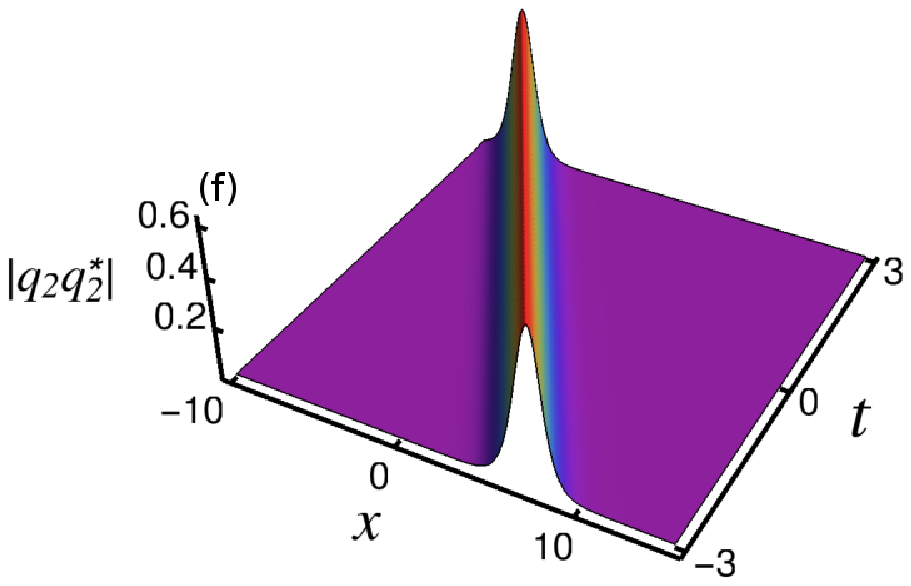}
\caption{(a) and (d) represent the absolute value of the  nonlinear Schr\"{o}dinger fields $|q_j(x,t)|$ drawn for the parameter values $k_1=1+i$, $\bar{k}_1=-1.4+i$, $\al_1^{(1)}=1+i$, $\al_1^{(2)}=1.5+i$, $\ba_1^{(1)}=1-i$ and $\ba_1^{(2)}=1-i$. (b) and (e) denote the absolute value of the fields $|q_j^*(-x,t)|$ plotted for the same values. (c) and (f) denote the $|q_j(x,t)q_j^*(-x,t)|$, $j=1,2$. Here, $*$ in $q_j^*$ in the present figures and the subsequent figures represent the fields $q_j^*(-x,t)$.}
\label{fig1}
\end{figure}
 The one soliton solution (\ref{2.14a})-(\ref{2.14b}) can also be rewritten as
\bes
\bea
q_j(x,t)=\frac{A_j(k_1+\bar{k}_1)\e^{\frac{(\bar{\xi}_{1R}-\xi_{1R})}{2}+i\frac{(\bar{\xi}_{1I}-\xi_{1I})}{2}}}{2i[\cosh(\chi_1)\cos(\chi_2)+i\sinh(\chi_1)\sin(\chi_2)]},\label{2.15a}
\eea  
and
\bea
q^*_j(-x,t)=\frac{\hat{A}_j(k_1+\bar{k}_1)\e^{\frac{-(\bar{\xi}_{1R}-\xi_{1R})}{2}-i\frac{(\bar{\xi}_{1I}-\xi_{1I})}{2}}}{2i[\cosh(\chi_1)\cos(\chi_2)+i\sinh(\chi_1)\sin(\chi_2)]},\label{2.15b}
\eea
respectively. In the above, the complex coefficients
\bea
\hspace{-1.0cm}A_j=\frac{\al_1^{(j)}}{\sqrt{(\al^{(1)}_1\ba^{(1)}_1+\al^{(2)}_1\ba^{(2)}_1)}},~\hat{A}_j=\frac{\ba_1^{(j)}}{\sqrt{(\al^{(1)}_1\ba^{(1)}_1+\al^{(2)}_1\ba^{(2)}_1)}},~ j=1,2\label{2.15c},
\eea \ees
and $\chi_1= \frac{\bar{\xi}_{1R}+\xi_{1R}+\Del_R}{2}$, $\chi_2=\frac{\bar{\xi}_{1I}+\xi_{1I}+\Del_I}{2}$, $\xi_{1I}=k_{1R}x+(-k_{1I}^{2}+k_{1R}^{2})t$, $\bar{\xi}_{1I}=\bar{k}_{1R}x+(-\bar{k}_{1R}^{2}+\bar{k}_{1I}^{2})t$, $\xi_{1R}=-k_{1I}(x+2k_{1R}t)$, $\bar{\xi}_{1R}=-\bar{k}_{1I}(x-2\bar{k}_{1R}t)$, $\Del_R=\frac{1}{2}\log{\bigg(\frac{|\al_1^{(1)}\ba_1^{(1)}+\al_1^{(2)}\ba_1^{(2)}|^2}{|k_1+\bar{k}_1|^2}\bigg)}$ and $\Del_I=\frac{-1}{2}\log\bigg(\frac{(\al_1^{(1)}\ba_1^{(1)}+\al_1^{(2)}\ba_1^{(2)})(k_1^{*}+\bar{k}_1^*)^2}{(\al_1^{(1*)}\ba_1^{(1*)}+\al_1^{(2*)}\ba_1^{(2*))}(k_1+\bar{k}_1)^2}\bigg)$. Here, $k_{1R}$ and $k_{1I}$, $\bar{k}_{1R}$ and $\bar{k}_{1I}$ are the real and imaginary parts of the wave numbers $k_1$ and $\bar{k}_1$, respectively. Similarly, $\xi_{1R}$ and $\xi_{1I}$, $\bar{\xi}_{1R}$ and $\bar{\xi}_{1I}$ are the real and imaginary parts of the wave numbers $\xi_1$ and $\bar{\xi}_1$, respectively.  To the best of our knowledge the one bright soliton solution given above is more general than the one already reported in the literature \cite{8a}. 

\subsection{Some remarkable features of degenerate nonlocal soliton}
For Eq. (\ref{1.1a}), we define the quasi-intensity (quasi-power) of solitons in both the modes as \cite{3a,3}
\ben
I_j=A_j\hat{A}_j, ~~j=1,2 \label{I}.
\een
In the local case the intensity of soliton is usually calculated by taking absolute squares of polarization vectors of the nonlinear Schr\"{o}dinger field whereas in the nonlocal case the intensity is calculated by multiplying the polarization vectors nonlinear Schr\"{o}dinger fields $q_j(x,t)$ by the polarization vectors of fields $q_j^*(-x,t)$. Here, hat in $\hat{A}_j$, $j=1,2$, denotes the polarization vectors present in field (\ref{2.15b}). 

 Using the expression (\ref{I}), a conserved quantity can be brought out in terms of the polarization vectors of the solitons of both the components, that is
\ben
\mathcal{I}=\int_{-\infty}^{\infty}(q_1(x,t)q_1^*(-x,t)+(q_2(x,t)q_2^*(-x,t))dx. 
\een As far as the one-soliton solution, the above form yields 
\ben
A_1\cdot \hat{A}_1+A_2\cdot\hat{A}_2=1 \label{II},
\een
where $A_1$, $A_2$, $\hat{A}_1$ and $\hat{A}_2$ are defined in Eq. (\ref{2.15c}).

We note here that for a specific parametric choice the one-soliton solution Eq. (\ref{2.15a})-(\ref{2.15b}) admits singularities for finite values of $t$ at $x=0$ when the following condition is satisfied:
\ben
\Del_R(k_{1R}^2-k_{1I}^2+\bar{k}_{1I}^2-\bar{k}_{1R}^2)=2[(2n+1)\pi-\Del_{I}](k_{1R}k_{1I}+\bar{k}_{1R}\bar{k}_{1I}),
\label{s}
\een
where, $n=0,1,2...$, respectively.  

 The long time evolution of the degenerate one soliton solution brings out yet another interesting feature for the nonlocal case. In  Fig. 1 we plot the absolute value of the degenerate one soliton solution $q_j(x,t)$ given in (\ref{2.14a}) for the parametric values $k_1=1+i$, $\bar{k}_1=-1.4+i$, $\al_1^{(1)}=1+i$, $\al_1^{(2)}=1.5+i$, $\ba_1^{(1)}=1-i$, $\ba_1^{(2)}=1-i$. As one can see in Figs. 1a and 1d, the amplitudes of the soliton in both the modes decay as $t\rightarrow+\infty$ in the $-x$ direction. The absolute value of the fields $q_j^*(-x,t)$ grow at $t\rightarrow+\infty$ in the $-x$ direction which is illustrated in Figs. 1b and 1e for the same parametric values. In other words a simultaneous loss and gain occur in the amplitudes of the solitons in the modes $q_j$ and $q_j^*$, $j=1,2$. However, a stable propagation of soliton can be visualized in the case  $|q_j(x,t)q_j^{*}(-x,t)|$ which is demonstrated in Figs. 1c and 1f. The amplitudes (or energy) of the soliton are preserved as specified by the conserved quantity of Eq. (\ref{1.1a}). In view of $\cal{PT}$-symmetric classical optics, the real and imaginary parts of the $\cal{PT}$-symmetric self induced potential,  $V(x,t)=\sum_{j=1}^{2}q_j(x,t)q_j^*(-x,t)$, that is present in the system (\ref{1.1a}). The stable propagation occurs due to the combined effect of loss and gain. 

 The complex amplitudes of the soliton in both the modes are $\frac{A_j(k_1+\bar{k}_1)}{2i}$,  $j=1,2$ where $A_j$'s are unit polarization vectors which are given in (\ref{2.15c}) . The soliton in the first and second components travels with the same velocity, that is $\frac{2(k_{1R}k_{1I}-\bar{k}_{1R}\bar{k}_{1I})}{(k_{1I}+\bar{k}_{1I})}$. We call such soliton as degenerate  soliton. The central position of the degenerate soliton in the two modes given by  $\frac{\Del_{R}}{(k_{1I}+\bar{k}_{1I})}=\frac{1}{(k_{1I}+\bar{k}_{1I})}\ln\frac{|\Gamma_{11}|}{|\kappa_{11}|}$. We recall here that in the local case, the velocity of the soliton is represented by the imaginary part of the wavenumbers \cite{6a,6b,7a}. 
\subsection{Sub-cases of general soliton solution}
 From the one-bright soliton solution, (\ref{2.14a}), we can also extract a two parameter family of breathing one-soliton solution which is reported in Ref. \cite{8a} by considering $\al^{(j)}_1=-\sqrt{2}(\eta_1+\bar{\eta_1})\e^{i\bar{\theta}_j}$, $\ba^{(j)}_1=-\sqrt{2}(\eta_1+\bar{\eta_1})\e^{i\theta_j}$, $j=1,2$, where $\theta_j$, $\bar{\theta}_j$, $\eta_1$ and $\bar{\eta_1}$ are all real parameters and by restricting the wave numbers $k_1$ and $\bar{k}_1$ as $k_1=2i\eta_1$ and $\bar{k}_1=2i\bar{\eta_1}$ (pure imaginary). Substituting these  restrictions in Eq. (\ref{2.14a}), we obtain
\bes
\ben
q_j(x,t)=-\frac{\sqrt{2}(\eta_1+\bar{\eta_1})\e^{i\bar{\theta}_j}e^{4i\bar{\eta_1}^{2}t}\e^{-2\bar{\eta_1}x}}{1+\e^{i(\theta_1+\bar{\theta}_1)}\e^{4i(\bar{\eta_1}^{2}-\eta_1^{2})t}\e^{-2(\eta_1+\bar{\eta_1})x}}, ~~j=1,2.
\label{2.16a}
\een
The solution (\ref{2.16a}) coincides with the one reported in Ref. \cite{8a}.   

 We can also obtain a similar expression for fields $q^{*}_j(-x,t)$ by imposing the same restrictions on Eq. (\ref{2.14b}). Doing so, we obtain
\ben
q^{*}_j(-x,t)=-\frac{\sqrt{2}(\eta_1+\bar{\eta_1})\e^{i\theta_j}\e^{-4i\eta_1^{2}t}\e^{-2\eta_1 x}}{1+\e^{i(\theta_1+\bar{\theta}_1)}\e^{4i(\bar{\eta_1}^{2}-\eta_1^{2})t}\e^{-2(\eta_1+\bar{\eta_1})x}}, ~j=1,2. \label{2.16b}
\een  

The above two parameter solutions develop a singularity in finite time which may be verified from the condition given in Eq. (\ref{s}) along with $k_{1R}=\bar{k}_{1R}=0$, $k_{1I}=2\eta_1$, $\bar{k}_{1I}=2\bar{\eta_1}$, $\al^{(j)}_1=-\sqrt{2}(\eta_1+\bar{\eta_1})\e^{i\bar{\theta}_j}$, $\ba^{(j)}_1=-\sqrt{2}(\eta_1+\bar{\eta_1})\e^{i\theta_j}$, $j=1,2$, and $\theta_1+\bar{\theta}_1=\theta_2+\bar{\theta}_2$. 
From the expressions (\ref{2.16a}) and (\ref{2.16b}) it is noted that $q^{*}_j(-x,t)$ is parity conjugate of $q_j(x,t)$.

 We can capture the envelop soliton solution of the local Manakov equation, that is
\ben
q_j(x,t)=-\sqrt{2}\eta\e^{-i(\theta_j-4\eta^{2}t)}\sech(2\eta x),
\een
 \ees
 by considering $\eta_1=\bar{\eta_1}=\eta$ and $\theta_j=-\bar{\theta}_j$, $j=1,2$, and imposing the above restrictions in the soliton solution (\ref{2.14a}). 
\section{Degenerate two bright soliton solution}
We obtain the degenerate two-soliton solution of (\ref{1.1a}) is, 
\bes \bea
q_j(x,t)&=&\frac{\al_{1}^{(j)}\e^{\bar{\xi}_1}+\al_{2}^{(j)}\e^{\bar{\xi}_2}+\e^{\xi_1+\bar{\xi}_1+\bar{\xi}_2+\Del_1^{(j)}}+\e^{\xi_2+\bar{\xi}_1+\bar{\xi}_2+\Del_2^{(j)}}}{D}, \label{21a}\\
q_j^*(-x,t)&=&\frac{\ba_{1}^{(j)}\e^{\xi_{1}}+\ba_{2}^{(j)}\e^{\xi_{2}}+\e^{\bar{\xi}_1+\xi_1+\xi_2+\ga_1^{(j)}}+\e^{\bar{\xi}_2+\xi_1+\xi_2+\ga_2^{(j)}}}{D},\label{21b}\\
D&=&1+\e^{\xi_{1}+\bar{\xi_1}+\del_1}+\e^{\xi_{1}+\bar{\xi_2}+\del_2}+\e^{\xi_{2}+\bar{\xi_1}+\del_3}+\e^{\xi_{2}+\bar{\xi_2}+\del_4}\nonumber\\&&+\e^{\xi_{1}+\bar{\xi_1}+\xi_{2}+\bar{\xi_2}+\del_5},\nonumber
\eea\ees
where $\xi_{j}=i k_{j}x-ik_{j}^{2}t$, $\bar{\xi}_j=i \bar{k}_{j}x+i\bar{k}_{j}^{2}t$, $j=1,2$ and the other constants are given in Appendix B. We note here that the above degenerate two-soliton solution is obtained from the expressions $(\ref{3.1})-(\ref{3.3})$ given in Appendix C, after appropriate factorization.  This is because the expression for the functions $f(x,t)$ and $f^*(-x,t)$ for degenerate one-soliton solution as well as two-soliton solution are equal at all orders of $\epsilon$. Due to this fact the degenerate two-soliton solution (\ref{3.1})-(\ref{3.3}) gets factorized into the above simple form. One can easily verified that expressions (\ref{21a})-(\ref{21b}) satisfy Eqs. (\ref{1.1a}) and   (\ref{1.1b}) simultaneously.  The above degenerate two bright soliton solution is characterized by twelve complex parameters, namely $\al_1^{(j)}$, $\al_2^{(j)}$, $\ba_1^{(j)}$, $\ba_2^{(j)}$, $k_j$ and $\bar{k}_j$, $j=1,2$.In the second part of the present work, we discuss the interaction between the degenerate two solitons in detail by carefully examining the two-soliton solution in the asymptotic regime.    

\section{Conclusion}
In this work,  we have constructed  more general  one and two soliton solutions for the nonlocal Manakov equation through a nonstandard bilinearization procedure. The obtained one- and two-soliton solutions are more general than the already reported ones. Besides deriving the soliton solutions, we have discussed the special features of the obtained soliton solutions. Next, we plan to investigate the collision dynamics through intensity redistribution, phase shift and relative separation distance by performing the asymptotic analysis of the two soliton solutions reported in this paper. 

{\bf \section*{Acknowledgements}}
The work of MS forms part of a research project sponsored by DST-SERB, Government of India, under the Grant No. EMR/2016/001818. The research work of ML is supported by a SERB Distinguished Fellowship and also forms part of the DAE-NBHM research project (2/48 (5)/2015/NBHM (R.P.)/R\&D-II/14127).

{\bf \section*{Appendix}}
\section*{A. The constants appear in non-degenerate one-soliton solution (\ref{nod1})-(\ref{nod2}) }
The constants which appear in the non-degenerate one-soliton solution (\ref{nod1})-(\ref{nod2}) have the explicit forms,
\bea
&&\e^{\Del_1^{(j)}}=\frac{(-1)^j(\bar{k}_1^{(1)}-\bar{k}_1^{(2)})\al_1^{(1)}\al_1^{(2)}\ba_1^{(3-j)}}{(\bar{k}_1^{(j)}+k_1^{(3-j)})(k_1^{(3-j)}+\bar{k}_1^{(3-j)})^2},\\
&&\e^{\ga_1^{(j)}}=\frac{(-1)^j(k_1^{(1)}-k_1^{(2)})\al_1^{(3-j)}\ba_1^{(1)}\ba_1^{(2)}}{(k_1^{(j)}+\bar{k}_1^{(3-j)})(k_1^{(3-j)}+\bar{k}_1^{(3-j)})^2}, ~ j=1,2,\\
&&\e^{\del_1}=\frac{-\al_1^{(1)}\ba_1^{(1)}}{(k_1^{(1)}+\bar{k}_1^{(1)})^2},~\e^{\del_2}=\frac{-\al_1^{(2)}\ba_1^{(2)}}{(k_1^{(2)}+\bar{k}_1^{(2)})^2},\\
&&\e^{\del_3}=\frac{\al_1^{(1)}\al_1^{(2)}\ba_1^{(1)}\ba_1^{(2)}(k_1^{(1)}-k_1^{(2)})(\bar{k}_1^{(1)}-\bar{k}_1^{(2)})}{(k_1^{(1)}+\bar{k}_1^{(1)})^2(\bar{k}_1^{(1)}+k_1^{(2)})(k_1^{(1)}+\bar{k}_1^{(2)})(k_1^{(2)}+\bar{k}_1^{(2)})^2}.
\eea 
\section*{B. The constants which appear in the reduced form of two-soliton solution (\ref{21a})-(\ref{21b}) }
The following constants appear in the two-soliton solution (\ref{21a})-(\ref{21b})  
\bes\bea
&&\e^{\Del_1^{(j)}}=\bar{\varrho}_{12}[(-1)^{3-j}k_1\ba_1^{(3-j)}\nu_1+\bar{k}_2\al_2^{(j)}\Gamma_{11}-\bar{k}_1\al_1^{(j)}\Gamma_{21}]/\kappa_{11}\kappa_{12},\\
&&\e^{\Del_2^{(j)}}=\bar{\varrho}_{12}[(-1)^{3-j}k_2\ba_2^{(3-j)}\nu_1+\bar{k}_2\al_2^{(j)}\Gamma_{12}-\bar{k}_1\al_1^{(j)}\Gamma_{22}]/\kappa_{21}\kappa_{22},\\
&&\e^{\ga_1^{(j)}}=\varrho_{12}[k_2\ba_2^{(j)}\Gamma_{11}-k_1\ba_1^{(j)}\Gamma_{12}+(-1)^{(j)}\bar{k}_1\al_1^{(3-j)}\nu_2]/\kappa_{11}\kappa_{21},\\
&&\e^{\ga_2^{(j)}}=\varrho_{12}[k_2\ba_2^{(j)}\Gamma_{21}-k_1\ba_1^{(j)}\Gamma_{22}+(-1)^{(j)}\bar{k}_2\al_2^{(3-j)}\nu_2]/\kappa_{12}\kappa_{22},\\
&&\e^{\del_{1}}=-\frac{\Gamma_{11}}{\kappa_{11}},~\e^{\del_{2}}=-\frac{\Gamma_{21}}{\kappa_{12}},~\e^{\del_{3}}=-\frac{\Gamma_{12}}{\kappa_{21}},~\e^{\del_{4}}=-\frac{\Gamma_{22}}{\kappa_{22}}, \\
&&\e^{\del_{5}}=\varrho_{12}\bar{\varrho}_{12}(\varrho_{1}\Gamma_{11}\Gamma_{22}-\varrho_{2}\nu_1\nu_2-\varrho_{3}\Gamma_{21}\Gamma_{12})/\kappa_{11}\kappa_{12}\kappa_{21}\kappa_{22}.
\eea\ees

\section*{C. An un-factored degenerate two-soliton solution}
A general un-factored degenerate two-soliton solution can be deduced by considering the following forms of seed solution for the functions  $g_1^{(j)}(x,t)$ and $g_1^{(j)*}(-x,t)$, for Eq. (\ref{lo}) that is
\bes 
\bea
&&g^{(j)}_1(x,t)=\al_{1}^{(j)}\e^{\bar{\xi}_1}+\al_{2}^{(j)}\e^{\bar{\xi}_2},~\bar{\xi_j}=i \bar{k_{j}}x+i\bar{k_{j}^{2}}t \label{2s.a},\\
&&g^{(j)*}_1(-x,t)=\ba_{1}^{(j)}\e^{\xi_1}+\ba_{2}^{(j)}\e^{\xi_2}, \xi_{j}=i k_{j}x-ik_{j}^{2}t, ~j=1,2 \label{2s.b}.
\eea \ees
The above form of seed solutions truncates the series expansions (\ref{2.4a})-(\ref{2.4c}) at in $7$-th order in $g^{(j)}(x,t)$ and $g^{(j)*}(-x,t)$, at $8$-th order in $f(x,t)$ and $f^{*}(-x,t)$ and $6$-th order in $s^{(1)}(-x,t)$ and $s^{(2)}(-x,t)$.  By solving the resultant equations that are arise at each order of $\epsilon$, we have obtained the following expressions  for the unknown functions $g^{(j)}(x,t)$, $g^{(j)*}(-x,t)$ and $f(x,t)$,
\bes\bea
g^{(j)}(x,t)&=&\al_{1}^{(j)}\e^{\bar{\xi}_1}+\al_{2}^{(j)}\e^{\bar{\xi}_2}+\e^{\xi_{1}+2\bar{\xi}_1+\Del_{1}^{(j)}}+\e^{\xi_{2}+2\bar{\xi_1}+\Del_{2}^{(j)}}+\e^{\xi_{1}+2\bar{\xi_2}+\Del_{3}^{(j)}}\nonumber \\
&&+\e^{\xi_{2}+2\bar{\xi_2}+\Del_{4}^{(j)}} +\e^{\xi_{1}+\bar{\xi_1}+\bar{\xi_2}+\Del_{5}^{(j)}}+\e^{\xi_{2}+\bar{\xi_1}+\bar{\xi_2}+\Del_{6}^{(j)}}+\e^{2\xi_{1}+2\bar{\xi_1}+\bar{\xi_2}+\Del_{7}^{(j)}}\nonumber\\
&&+\e^{2\xi_{1}+\bar{\xi_1}+2\bar{\xi_2}+\Del_{8}^{(j)}}+\e^{2\xi_{2}+2\bar{\xi_1}+\bar{\xi_2}+\Del_{9}^{(j)}}+\e^{2\xi_{2}+\bar{\xi_1}+2\bar{\xi_2}+\mu_{1}^{(j)}}\nonumber\\
&&+\e^{\xi_{1}+\bar{\xi_1}+\xi_2+2\bar{\xi_2}+\mu_{2}^{(j)}}+\e^{\xi_{1}+2\bar{\xi_1}+\xi_2+\bar{\xi_2}+\mu_{3}^{(j)}}+\e^{2\xi_{1}+2\bar{\xi_1}+\xi_{2}+2\bar{\xi_2}+\mu_{4}^{(j)}}\nonumber\\
&&+\e^{2\bar{\xi_1}+2\bar{\xi_2}+\xi_{1}+2\xi_{2}+\mu_{5}^{(j)}}\label{3.1}\\
g^{(j)*}(-x,t)&=&\ba_{1}^{(j)}\e^{\xi_{1}}+\ba_{2}^{(j)}\e^{\xi_{2}}+\e^{2\xi_{1}+\bar{\xi_1}+\ga_{1}^{(j)}}+\e^{2\xi_{1}+\bar{\xi_2}+\ga_{2}^{(j)}}+\e^{2\xi_{2}+\bar{\xi_1}+\ga_{3}^{(j)}}\nonumber \\ 
&&+\e^{2\xi_{2}+\bar{\xi_2}+\ga_{4}^{(j)}}+\e^{\xi_{1}+\bar{\xi_1}+\xi_{2}+\ga_{5}^{(j)}}+\e^{\xi_{1}+\xi_{2}+\bar{\xi_2}+\ga_{6}^{(j)}}+\e^{2\xi_{1}+2\bar{\xi_1}+\xi_{2}+\ga_{7}^{(j)}}\nonumber\\
&&+\e^{\xi_{1}+2\bar{\xi_1}+2\xi_{2}+\ga_{8}^{(j)}}+\e^{2\xi_{1}+2\bar{\xi_2}+\xi_{2}+\ga_{9}^{(j)}}+\e^{2\xi_{2}+2\bar{\xi_2}+\xi_{1}+\varphi_{1}^{(j)}}\nonumber\\
&&+\e^{2\xi_{1}+\bar{\xi_1}+\xi_{2}+\bar{\xi_2}+\varphi_{2}^{(j)}}+\e^{\xi_{1}+\bar{\xi_1}+2\xi_{2}+\bar{\xi_2}+\varphi_{3}^{(j)}}+\e^{2\xi_{1}+2\bar{\xi_1}+2\xi_{2}+\bar{\xi_2}+\varphi_{4}^{(j)}}\nonumber\\
&&+\e^{2\xi_{1}+\bar{\xi_1}+2\xi_{2}+2\bar{\xi_2}+\varphi_{5}^{(j)}}\label{3.2}\\
f(x,t)&=&1+\e^{\xi_{1}+\bar{\xi_1}+\del_1}+\e^{\xi_{2}+\bar{\xi_1}+\del_2}+\e^{\xi_{1}+\bar{\xi_2}+\del_3}+\e^{\xi_{2}+\bar{\xi_2}+\del_4}+\e^{2(\xi_{1}+\bar{\xi_1})+\del_{11}}\nonumber\\ 
&&+\e^{2(\xi_{2}+\bar{\xi_1})+\del_{12}}+\e^{2(\xi_{1}+\bar{\xi_2})+\del_{13}}+\e^{2(\xi_{2}+\bar{\xi_2})+\del_{14}}+\e^{2\bar{\xi_1}+\xi_{1}+\xi_{2}+\del_{15}}\nonumber\\
&&+\e^{2\bar{\xi_2}+\xi_{1}+\xi_{2}+\del_{16}}+\e^{2\xi_{1}+\bar{\xi_1}+\bar{\xi_2}+\del_{17}}+\e^{2\xi_{2}+\bar{\xi_1}+\bar{\xi_2}+\del_{18}}+\e^{\xi_{1}+\bar{\xi_1}+\xi_{2}+\bar{\xi_2}+\del_{19}}\nonumber\\
&&+\e^{2\xi_{1}+2\bar{\xi_1}+\xi_{2}+\bar{\xi_2}+\del_{21}}+\e^{2\xi_{1}+\bar{\xi_1}+\xi_{2}+2\bar{\xi_2}+\del_{22}}+\e^{\xi_{1}+2\bar{\xi_1}+2\xi_{2}+\bar{\xi_2}+\del_{23}}\nonumber\\
&&+\e^{\xi_{1}+\bar{\xi_1}+2\xi_{2}+2\bar{\xi_2}+\del_{24}}+\e^{2(\xi_{1}+\bar{\xi_1}+\xi_{2}+\bar{\xi_2})+\del_{31}}\equiv f^{*}(-x,t).
\label{3.3}
\eea
\ees 
The explicit expression of all the constants that appear in  two-soliton solution are given as
\bea
&&\e^{\del_{1}}=-2\frac{\Gamma_{11}}{\kappa_{11}},~\e^{\del_{2}}=-2\frac{\Gamma_{12}}{\kappa_{21}},~\e^{\del_{3}}=-2\frac{\Gamma_{21}}{\kappa_{12}},~\e^{\del_{4}}=-2\frac{\Gamma_{22}}{\kappa_{22}},\nonumber\\
&&\Gamma_{11}=(\al^{(1)}_1\ba^{(1)}_1+\al^{(2)}_1\ba^{(2)}_1), \Gamma_{12}=(\al^{(1)}_1\ba^{(1)}_2+\al^{(2)}_1\ba^{(2)}_2),\nonumber\\
&&\Gamma_{21}=(\al^{(1)}_2\ba^{(1)}_1+\al^{(2)}_2\ba^{(2)}_1), \Gamma_{22}=(\al^{(1)}_2\ba^{(1)}_2+\al^{(2)}_2\ba^{(2)}_2),~ \kappa_{lm}=(k_l+\bar{k}_m)^2 ,~l,m=1,2.\nonumber\\
&&\e^{\Del_{1}^{(j)}}=-\frac{\al_1^{(j)}\Gamma_{11}}{\kappa_{11}},~\e^{\Del_{2}^{(j)}}=-\frac{\al_1^{(j)}\Gamma_{12}}{\kappa_{21}},~\e^{\Del_{3}^{(j)}}=-\frac{\al_2^{(j)}\Gamma_{21}}{\kappa_{12}},~\e^{\Del_{4}^{(j)}}=-\frac{\al_2^{(j)}\Gamma_{22}}{\kappa_{22}},\nonumber \eea\bea
&&\e^{\Del_{5}^{(j)}}=-\bigg(\al_1^{(j)}\Gamma_{21}(k_1+\bar{k}_1)(k_1+2\bar{k}_1-\bar{k}_2)+\al_2^{(j)}\Gamma_{11}(k_1+\bar{k}_2)(k_1+2\bar{k}_2-\bar{k}_1)\bigg)/\kappa_{11}\kappa_{12},~~~~\nonumber\\
&&\e^{\Del_{6}^{(j)}}=-{\bigg(\al_1^{(j)}\Gamma_{22}(k_2+\bar{k}_1)(k_2+2\bar{k}_1-\bar{k}_2)+\al_2^{(j)}\Gamma_{12}(k_2+\bar{k}_2)(k_2+2\bar{k}_2-\bar{k}_1)\bigg)/ \kappa_{21}\kappa_{22}},~~~\nonumber\\
&&\e^{\ga_{1}^{(j)}}=-\frac{\ba_1^{(j)}\Gamma_{11}}{\kappa_{11}},~\e^{\ga_{2}^{(j)}}=-\frac{\ba_1^{(j)}\Gamma_{21}}{\kappa_{12}},~\e^{\ga_{3}^{(j)}}=-\frac{\ba_2^{(j)}\Gamma_{12}}{\kappa_{21}},~\e^{\ga_{4}^{(j)}}=-\frac{\ba_2^{(j)}\Gamma_{22}}{\kappa_{22}},\nonumber\\
&&\e^{\ga_{5}^{(j)}}=-{\bigg(\ba_1^{(j)}\Gamma_{12}(k_1+\bar{k}_1)(\bar{k}_1+2k_1-k_2)+\ba_2^{(j)}\Gamma_{11}(k_2+\bar{k}_1)(\bar{k}_1+2k_2-k_1)\bigg)/\kappa_{11}\kappa_{21}},~~~\nonumber\\
&&\e^{\ga_{6}^{(j)}}=-{\bigg(\ba_1^{(j)}\Gamma_{22}(k_1+\bar{k}_2)(\bar{k}_2+2k_1-k_2)+\ba_2^{(j)}\Gamma_{21}(k_2+\bar{k}_2)(\bar{k}_2+2k_2-k_1)\bigg)/\kappa_{12}\kappa_{22}}.~~~\nonumber\\
&&\e^{\del_{11}}=\frac{\Gamma_{11}^2}{\kappa_{11}^2},~\e^{\del_{12}}=\frac{\Gamma_{12}^2}{\kappa_{21}^2},~\e^{\del_{13}}=\frac{\Gamma_{21}^2}{\kappa_{12}^2},~\e^{\del_{14}}=\frac{\Gamma_{22}^2}{\kappa_{22}^2},~\e^{\del_{15}}=\frac{2\Gamma_{11}\Gamma_{12}}{\kappa_{11}\kappa_{21}},\nonumber\\
&&\e^{\del_{16}}=\frac{2\Gamma_{21}\Gamma_{22}}{\kappa_{12}\kappa_{22}},~\e^{\del_{17}}=\frac{2\Gamma_{11}\Gamma_{21}}{\kappa_{11}\kappa_{12}},~\e^{\del_{18}}=\frac{2\Gamma_{12}\Gamma_{22}}{\kappa_{21}\kappa_{22}},\nonumber\\
&&\e^{\del_{19}}=\frac{2(\kappa_{21}\kappa_{12})^{\frac{1}{2}}\Lambda_3+2(\kappa_{11}\kappa_{22})^{\frac{1}{2}}\Lambda_4+4\Lambda_5}{\kappa_{11}\kappa_{12}\kappa_{21}\kappa_{22}},\nonumber\\
&&\Lambda_3=(k_1(2\bar{k}_1+k_2-\bar{k}_2)+2k_2\bar{k}_2+\bar{k}_1(\bar{k}_2-k_2))(\al_1^{(1)}\ba_1^{(1)}\al_2^{(2)}\ba_2^{(2)}+\al_1^{(1)}\ba_1^{(2)}\al_2^{(1)}\ba_2^{(1)}),~~\nonumber\\
&&\Lambda_4=(-k_2\bar{k}_2+\bar{k}_1(2k_2+\bar{k}_2)+k_1(2\bar{k}_2-\bar{k}_1+k_2))(\al_1^{(2)}\ba_1^{(1)}\al_2^{(1)}\ba_2^{(2)}+\al_1^{(1)}\ba_1^{(2)}\al_2^{(2)}\ba_2^{(1)}),~~\nonumber\\
&&\Lambda_5=(k_2^2\bar{k}_2^2+k_2\bar{k}_1\bar{k}_2(\bar{k}_2-k_2)+\bar{k}_1^2(k_2^2+k_2\bar{k}_2+\bar{k}_2^2)+k_1^2(\bar{k}_1^2+k_2^2+\bar{k}_1(k_2-\bar{k}_2)+k_2\bar{k}_2\nonumber\\
&&~~~~~~~+\bar{k}_2^2)+k_1[k_2\bar{k}_2(k_2-\bar{k}_2)+\bar{k}_1^2(\bar{k}_2-k_2)+\bar{k}_1(k_2^2+5k_2\bar{k}_2+\bar{k}_2^2)])\nonumber\\
&&~~~~~~~~(\al_1^{(2)}\ba_1^{(2)}\al_2^{(2)}\ba_2^{(2)}+\al_1^{(1)}\ba_1^{(1)}\al_2^{(1)}\ba_2^{(1)}).\nonumber\\
&&\e^{\Del_{7}^{(j)}}={\bar{\varrho}_{12}\Gamma_{11}\bigg((-1)^{j}k_1\ba_1^{(3-j)}\nu_1-\bar{k}_2\al_2^{(j)}\Gamma_{11}+\bar{k}_1\al_1^{(j)}\Gamma_{21}\bigg)/\kappa_{11}^2\kappa_{12}},\nonumber\\
&&\e^{\Del_{8}^{(j)}}={\bar{\varrho}_{12}\Gamma_{21}\bigg((-1)^{j}k_1\ba_1^{(3-j)}\nu_1-\bar{k}_2\al_2^{(j)}\Gamma_{11}+\bar{k}_1\al_1^{(j)}\Gamma_{21}\bigg)/\kappa_{11}\kappa_{12}^2},\nonumber\\
&&\e^{\Del_{9}^{(j)}}={\bar{\varrho}_{12}\Gamma_{12}\bigg((-1)^{j}k_2\ba_2^{(3-j)}\nu_1-\bar{k}_2\al_2^{(j)}\Gamma_{12}+\bar{k}_1\al_1^{(j)}\Gamma_{22}\bigg)/\kappa_{21}^2\kappa_{22}},\nonumber\\
&&\e^{\mu_{1}^{(j)}}={\bar{\varrho}_{12}\Gamma_{22}\bigg((-1)^{j}k_2\ba_2^{(3-j)}\nu_1-\bar{k}_2\al_2^{(j)}\Gamma_{12}+\bar{k}_1\al_1^{(j)}\Gamma_{22}\bigg)/\kappa_{21}\kappa_{22}^2},\nonumber\\
&&\e^{\ga_{7}^{(j)}}={\varrho_{12}\Gamma_{11}\bigg(-k_2\ba_2^{(j)}\Gamma_{11}+k_1\ba_1^{(j)}\Gamma_{12}+(-1)^{(3-j)}\bar{k}_1\al_1^{(3-j)}\nu_2\bigg)/\kappa_{11}^2\kappa_{21}},\nonumber\\
&&\e^{\ga_{8}^{(j)}}={\varrho_{12}\Gamma_{12}\bigg(-k_2\ba_2^{(j)}\Gamma_{11}+k_1\ba_1^{(j)}\Gamma_{12}+(-1)^{(3-j)}\bar{k}_1\al_1^{(3-j)}\nu_2\bigg)/\kappa_{11}\kappa_{21}^2},\nonumber\\
&&\e^{\ga_{9}^{(j)}}={\varrho_{12}\Gamma_{21}\bigg(-k_2\ba_2^{(j)}\Gamma_{21}+k_1\ba_1^{(j)}\Gamma_{22}+(-1)^{(3-j)}\bar{k}_2\al_2^{(3-j)}\nu_2\bigg)/\kappa_{12}^2\kappa_{22}},\nonumber\\
&&\e^{\varphi_{1}^{(j)}}={\varrho_{12}\Gamma_{22}\bigg(-k_2\ba_2^{(j)}\Gamma_{21}+k_1\ba_1^{(j)}\Gamma_{22}+(-1)^{(3-j)}\bar{k}_2\al_2^{(3-j)}\nu_2\bigg)/\kappa_{12}\kappa_{22}^2},\nonumber\eea \bea
&&\e^{\mu_2^{(j)}}=\frac{\bar{\varrho}_{12}\Lambda_6}{\kappa_{11}\kappa_{12}\kappa_{21}\kappa_{22}},~\e^{\mu_3^{(j)}}=\frac{\bar{\varrho}_{12}\Lambda_7}{\kappa_{11}\kappa_{12}\kappa_{21}\kappa_{22}},~\e^{\varphi_2^{(j)}}=\frac{\varrho_{12}\Lambda_8}{\kappa_{11}\kappa_{12}\kappa_{21}\kappa_{22}},\nonumber\\
&&\e^{\varphi_3^{(j)}}=\frac{\varrho_{12}\Lambda_9}{\kappa_{11}\kappa_{12}\kappa_{21}\kappa_{22}},\nonumber\\
&&\Lambda_6=\bigg(-2k_2^2\bar{k}_2\al_2^{(j)}\Gamma_{11}\Gamma_{22}+2\bar{k}_1^3\al_1^{(j)}\Gamma_{21}\Gamma_{22}+k_1^2[-2\bar{k}_2\al_2^{(j)}\Gamma_{21}\Gamma_{12}+\bar{k}_1(\al_2^{(j)}\Gamma_{11}\nonumber\\
&&~~~~~~~~~+\al_1^{(j)}\Gamma_{21})\Gamma_{22}+k_2\nu_1(-\al_2^{(j)}\nu_2+\ba_2^{(3-j)}(-1)^j\Gamma_{21})]+\bar{k}_1k_2[k_2\Gamma_{21}(\al_1^{(j)}\Gamma_{22}+\al_2^{(j)}\nonumber\\
&&~~~~~~~~~~~\Gamma_{12})+\bar{k}_2\al_2^{(j)}(\al_1^{(1)}(-2\ba_1^{(1)}\Gamma_{22}-\al_2^{(2)}\nu_2)+\al_1^{(2)}(-2\ba_1^{(2)}\Gamma_{22}+\al_2^{(1)}\nu_2))]\nonumber\\
&&~~~~~~~~~~+\bar{k}_1^2[k_2\Gamma_{21}(2\al_1^{(j)}\Gamma_{22}+(-1)^j\ba_2^{(3-j)}\nu_1)-\bar{k}_2\al_2^{(j)}(\al_1^{(1)}(\ba_2^{(1)}\Gamma_{21}+\ba_1^{(1)}\Gamma_{22})+\al_1^{(2)}\nonumber\\
&&~~~~~~~~~~(\ba_1^{(2)}\Gamma_{22}+\ba_2^{(2)}\Gamma_{21}))]+k_1[\bar{k}_1^2\Gamma_{22}(2\al_1^{(j)}\Gamma_{21}+(-1)^j\ba_1^{(3-j)}\nu_1)+k_2(k_2\nu_1(-1)^j((-1)^j\nonumber\\
&&~~~~~~~~~~~\al_2^{(j)}\nu_2+\ba_1^{(3-j)}\Gamma_{22})+\bar{k}_2\al_2^{(j)}(\Gamma_{11}\Gamma_{22}+\Gamma_{21}\Gamma_{12}))+\bar{k}_1[\bar{k}_2\al_2^{(j)}\big(\al_1^{(1)}(-2\ba_2^{(1)}\Gamma_{21}\nonumber\\
&&~~~~~~~~~~+\al_2^{(2)}\nu_2)+\al_1^{(2)}(-\al_2^{(1)}\nu_2-2\ba_2^{(2)}\Gamma_{21})\big)+k_2\big(-3\al_1^{(3-j)}\al_2^{(j)}(\ba_2^{(3-j)}\Gamma_{21}+\ba_1^{(3-j)}\Gamma_{22})\nonumber\\
&&~~~~~~~~~~+\al_1^{(j)}\big(\al_2^{(3-j)}\ba_1^{(3-j)}\Gamma_{22}+\al_2^{(3-j)}\ba_2^{(3-j)}\Gamma_{21}+2\al_2^{(j)2}\ba_1^{(3-j)}\ba_2^{(3-j)}-2\al_2^{(j)2}\ba_1^{(j)}\ba_2^{(j)}\big)\big)]]\bigg),\nonumber\\
&&\Lambda_7=\bigg(-k_2\Gamma_{11}[2\bar{k}_2^2\al_2^{(j)}\Gamma_{12}+k_2\bar{k}_2(2\al_2^{(j)}\Gamma_{12}+(-1)^{(3-j)}\ba_2^{(3-j)}\nu_1)+k_2^2(\al_1^{(j)}\Gamma_{22}+\al_2^{(j)}\nonumber\\
&&~~~~~~~~~~~\Gamma_{12})]+k_1^2[-\bar{k}_2(\al_1^{(j)}\Gamma_{21}+\al_2^{(j)}\Gamma_{11})\Gamma_{12}+2\bar{k}_1\al_1^{(j)}\Gamma_{11}\Gamma_{22}+(-1)^jk_2\nu_1((-1)^{(3-j)}\al_1^{(j)}\nonumber\\
&&~~~~~~~~~~\nu_2+\ba_2^{(3-j)}\Gamma_{11})]+\bar{k}_1\al_1^{(j)}[2k_2^2\Gamma_{21}\Gamma_{12}+\bar{k}_2^2\big(\al_1^{(1)}(\ba_2^{(1)}\Gamma_{21}+\ba_1^{(1)}\Gamma_{22})+\al_1^{(2)}(\ba_1^{(2)}\Gamma_{22}\nonumber\\
&&~~~~~~~~~~+\ba_2^{(2)}\Gamma_{21})\big)+k_2\bar{k}_2\big(\al_1^{(1)}(2\ba_2^{(1)}\Gamma_{21}-\al_2^{(2)}\nu_2)+\al_1^{(2)}(2\ba_2^{(2)}\Gamma_{21}+\al_2^{(1)}\nu_2)\big)]-k_1\nonumber\\
&&~~~~~~~~~~[\bar{k}_2^2(2\al_2^{(j)}\Gamma_{11}+(-1)^{(3-j)}\ba_1^{(3-j)}\nu_1)\Gamma_{12}+(-1)^{(3-j)}k_2^2\nu_1((-1)^j\al_1^{(j)}\nu_2+\ba_1^{(3-j)}\Gamma_{12})\nonumber\\
&&~~~~~~~~~~+k_2\bar{k}_2\big((-1)^{(3-j)}\ba_1^{(3-j)}\Gamma_{12}\nu_1+(-1)^{(3-j)}\ba_2^{(3-j)}\Gamma_{11}\nu_1-2\al_1^{(j)}\al_2^{(3-j)}(\ba_1^{(3-j)}\Gamma_{12}\nonumber\\
&&~~~~~~~~~~+\ba_2^{(3-j)}\Gamma_{11})+2\al_1^{(3-j)2}\al_2^{(j)}\ba_1^{(3-j)}\ba_{2}^{(3-j)}-2\al_1^{(j)2}\al_2^{(j)}\ba_1^{(j)}\ba_2^{(j)}\big)+\bar{k}_1\al_1^{(j)}\big(\bar{k}_2\nonumber\\
&&~~~~~~~~~~(-2\Gamma_{11}\Gamma_{22}+\nu_1\nu_2)+k_2(\Gamma_{11}\Gamma_{22}+\Gamma_{12}\Gamma_{21})\big)]\bigg),\nonumber\\
&&\Lambda_8=\bigg(-\bar{k}_1^2[\bar{k}_2(\ba_1^{(j)}\nu_1+(-1)^j\al_2^{(3-j)}\Gamma_{11})\nu_2+k_2\Gamma_{21}(\ba_2^{(j)}\Gamma_{11}+\ba_1^{(j)}\Gamma_{12})]-k_2\Gamma_{11}\nonumber\\
&&~~~~~~~~~~[2k_2^2\Gamma_{21}\ba_2^{(j)}+\bar{k}_2^2(\ba_2^{(j)}\Gamma_{21}+\ba_1^{(j)}\Gamma_{22})+k_2\bar{k}_2(2\ba_2^{(j)}\Gamma_{21}+(-1)^j\al_2^{(3-j)}\nu_2)]+\nonumber\\
&&~~~~~~~~~~k_1\ba_1^{(j)}[2\bar{k}_2^2\Gamma_{12}\Gamma_{21}+2\bar{k}_1^2\Gamma_{11}\Gamma_{22}+k_2\bar{k}_2(2\Gamma_{21}\Gamma_{12}+\nu_1\nu_2)+k_2^2(\Gamma_{12}\Gamma_{21}+\Gamma_{11}\Gamma_{22})\nonumber\\
&&~~~~~~~~~~+\bar{k}_1\big(-\bar{k}_2(\Gamma_{12}\Gamma_{21}+\Gamma_{11}\Gamma_{22})+k_2(2\Gamma_{11}\Gamma_{22}-\nu_2\nu_2)\big)]+\bar{k}_1[(-1)^{(3-j)}\bar{k}_2^2\nu_2\nonumber\\
&&~~~~~~~~~~((-1)^{(3-j)}\ba_1^{(j)}\nu_1+\al_1^{(3-j)}\Gamma_{21})-k_2^2\Gamma_{21}(2\ba_2^{(j)}\Gamma_{11}+(-1)^{(j)}\al_1^{(3-j)}\nu_2)+k_2\bar{k}_2\nonumber\\
&&~~~~~~~~~~\big(\al_1^{(j)}\ba_1^{(j)}(2\ba_1^{(j)}\Gamma_{22}+(-1)^{(3-j)}\al_2^{(3-j)}\nu_2)+\al_1^{(3-j)}((-1)^{(3-j)}\Gamma_{21}\nu_2+2\ba_1^{(j)}\ba_2^{(3-j)}\nonumber\\
&&~~~~~~~~~~\Gamma_{21}+3(-1)^{(3-j)}\al_2^{(3-j)}\ba_1^{(3-j)}\nu_2)\big)]\bigg),\nonumber\eea \bea
&&\Lambda_9=\bigg(-2k_2\bar{k}_2^2\Gamma_{11}\Gamma_{22}\ba_2^{(j)}+2k_1^3\ba_1^{(j)}\Gamma_{12}\Gamma_{22}+\bar{k}_1^2(-1)^j[(-1)^{(3-j)}2k_2\Gamma_{21}\ba_2^{(j)}+(-1)^{(3-j)}\nonumber\\
&&~~~~~~~~~\bar{k}_2\nu_2(\nu_1\ba_2^{(j)}+(-1)^{(j)}\al_2^{(3-j)}\Gamma_{12})]+k_1^2[\bar{k}_1\Gamma_{22}(2\ba_1^{(j)}\Gamma_{12}+(-1)^{(3-j)}\al_1^{(3-j)}\nu_2)+\bar{k}_2\Gamma_{12}\nonumber\\
&&~~~~~~~~~(2\ba_1^{(j)}\Gamma_{22}+(-1)^{(3-j)}\al_2^{(3-j)}\nu_2)-k_2\ba_2^{(j)}(\Gamma_{11}\Gamma_{22}+\Gamma_{12}\Gamma_{21})]+\bar{k}_1\bar{k}_2[(-1)^{(3-j)}\bar{k}_2\nonumber\\
&&~~~~~~~~~\nu_2(\al_1^{(3-j)}\Gamma_{22}+(-1)^{(3-j)}\ba_2^{(j)}\nu_1)+k_2\ba_2^{(j)}\big(\al_1^{(1)}(\ba_2^{(1)}\Gamma_{21}+\ba_1^{(1)}\Gamma_{22})+\al_1^{(2)}(\ba_1^{(2)}\Gamma_{22}\nonumber\\
&&~~~~~~~~~+\ba_2^{(2)}\Gamma_{21})\big)]+k_1[\bar{k}_1^2\Gamma_{22}(\ba_2^{(j)}\Gamma_{11}+\ba_1^{(j)}\Gamma_{12})+\bar{k}_2\big(\bar{k}_2\Gamma_{12}(\ba_1^{(j)}\Gamma_{22}+\ba_2^{(j)}\Gamma_{21})+k_2\ba_2^{(j)}\nonumber\\
&&~~~~~~~~~(-2\Gamma_{11}\Gamma_{22}+\nu_1\nu_2)\big)+\bar{k}_1\big(k_2\ba_2^{(j)}[\al_1^{(1)}(-2\ba_2^{(1)}\Gamma_{21}+\al_2^{(2)}\nu_2)+(-2\ba_2^{(2)}\Gamma_{21}-\al_2^{(1)}\nu_2)]\nonumber\\
&&~~~~~~~~~+\bar{k}_2[\al_1^{(j)}\ba_2^{(j)}(-2\ba_2^{(j)}\Gamma_{21}+(-1)^{(3-j)}\al_2^{(3-j)}\nu_2)+\al_1^{(3-j)}(-2\ba_1^{(j)}\ba_2^{(j)}\Gamma_{22}+(-1)^{(3-j)}\nu_2\nonumber\\
&&\Gamma_{22}+3(-1)^{(3-j)}\al_2^{(3-j)}\ba_2^{(j)}\nu_2)]\big)]\bigg),\nonumber\\
&&\nu_1=\al_1^{(2)}\al_2^{(1)}-\al_1^{(1)}\al_2^{(2)},~\nu_2=\ba_1^{(1)}\ba_2^{(2)}-\ba_1^{(2)}\ba_2^{(1)},~\varrho_{12}=(k_1-k_2),~\bar{\varrho}_{12}=(\bar{k}_1-\bar{k}_2),\nonumber \\
&&\e^{\del_{21}}={-2\varrho_{12}\bar{\varrho}_{12}\Gamma_{11}\bigg(\varrho_{1}\Gamma_{11}\Gamma_{22}-\varrho_{2}\nu_1\nu_2-\varrho_{3}\Gamma_{21}\Gamma_{12}\bigg)/\kappa_{11}^2\kappa_{12}\kappa_{21}\kappa_{22}},\nonumber\\
&&\e^{\del_{22}}={-2\varrho_{12}\bar{\varrho}_{12}\Gamma_{21}\bigg(\varrho_{1}\Gamma_{11}\Gamma_{22}-\varrho_{2}\nu_1\nu_2-\varrho_{3}\Gamma_{21}\Gamma_{12}\bigg)/\kappa_{11}\kappa_{12}^2\kappa_{21}\kappa_{22}},\nonumber\\
&&\e^{\del_{23}}={-2\varrho_{12}\bar{\varrho}_{12}\Gamma_{12}\bigg(\varrho_{1}\Gamma_{11}\Gamma_{22}-\varrho_{2}\nu_1\nu_2-\varrho_{3}\Gamma_{21}\Gamma_{12}\bigg)/\kappa_{11}\kappa_{12}\kappa_{21}^2\kappa_{22}},\nonumber\\
&&\e^{\del_{24}}={-2\varrho_{12}\bar{\varrho}_{12}\Gamma_{22}\bigg(\varrho_{1}\Gamma_{11}\Gamma_{22}-\varrho_{2}\nu_1\nu_2-\varrho_{3}\Gamma_{21}\Gamma_{12}\bigg)/\kappa_{11}\kappa_{12}\kappa_{21}\kappa_{22}^2},\nonumber\\
&&\varrho_{1}=(k_2\bar{k}_2+k_1\bar{k}_1),~\varrho_{2}=(k_1k_2+\bar{k}_1\bar{k}_2),~\varrho_{3}=(k_1\bar{k}_2+k_2\bar{k}_1).\nonumber
\\
&&\e^{\mu_{4}^{(j)}}=-\varrho_{12}\bar{\varrho}_{12}^2\bigg((-1)^{j}k_1\ba_1^{(3-j)}\nu_1-\bar{k}_2\al_2^{(j)}\Gamma_{11}+\bar{k}_1\al_1^{(j)}\Gamma_{21}\bigg)\nonumber\\
&&\hspace{1.2cm}\bigg(\varrho_{1}\Gamma_{11}\Gamma_{22}-\varrho_{2}\nu_1\nu_2-\varrho_{3}\Gamma_{21}\Gamma_{12}\bigg)/\kappa\kappa_{11}\kappa_{12}
,\nonumber\\
&&\e^{\mu_{5}^{(j)}}=-\varrho_{12}\bar{\varrho}_{12}^2\bigg((-1)^{j}k_2\ba_2^{(3-j)}\nu_1-\bar{k}_2\al_2^{(j)}\Gamma_{11}+\bar{k}_1\al_1^{(j)}\Gamma_{21}\bigg)\nonumber\\
&&\hspace{1.2cm}\bigg(\varrho_{1}\Gamma_{11}\Gamma_{22}-\varrho_{2}\nu_1\nu_2-\varrho_{3}\Gamma_{21}\Gamma_{12}\bigg)/\kappa\kappa_{21}\kappa_{22}
\nonumber\\
&&\e^{\varphi_{4}^{(j)}}=-\varrho_{12}^2\bar{\varrho}_{12}\bigg(-k_2\ba_2^{(j)}\Gamma_{11}+k_1\ba_1^{(j)}\Gamma_{12}+(-1)^{(3-j)}\bar{k}_1\al_1^{(3-j)}\nu_2\bigg)\nonumber\\
&&\hspace{1.2cm}\bigg(\varrho_{1}\Gamma_{11}\Gamma_{22}-\varrho_{2}\nu_1\nu_2-\varrho_{3}\Gamma_{21}\Gamma_{12}\bigg)/\kappa\kappa_{11}\kappa_{21},\nonumber\\
&&\e^{\varphi_{5}^{(j)}}=-\varrho_{12}^2\bar{\varrho}_{12}\bigg(-k_2\ba_2^{(j)}\Gamma_{21}+k_1\ba_1^{(j)}\Gamma_{22}+(-1)^{(3-j)}\bar{k}_2\al_2^{(3-j)}\nu_2\bigg)\nonumber\\
&&\hspace{1.2cm}\bigg(\varrho_{1}\Gamma_{11}\Gamma_{22}-\varrho_{2}\nu_1\nu_2-\varrho_{3}\Gamma_{21}\Gamma_{12}\bigg)/\kappa\kappa_{12}\kappa_{22},\nonumber\\
&&\e^{\del_{31}}=\varrho_{12}^2\bar{\varrho}_{12}^2\bigg(\varrho_{1}\Gamma_{11}\Gamma_{22}-\varrho_{2}\nu_1\nu_2-\varrho_{3}\Gamma_{21}\Gamma_{12}\bigg)^2/\kappa_{11}^2\kappa_{12}^2\kappa_{21}^2\kappa_{22}^2,\nonumber\\
&&\kappa=\kappa_{11}\kappa_{12}\kappa_{21}\kappa_{22}.\nonumber
\eea
We arrive the degenerate two-soliton solution by substituting the expression given in (\ref{3.1})-(\ref{3.3}) in Eq. (\ref{2.1}). The auxiliary functions are found to be
\bea
&&s^{(1)}(-x,t)=s^{(2)}(-x,t)=\Gamma_{11}\e^{\xi_1+\bar{\xi}_1}+\Gamma_{21}\e^{\xi_1+\bar{\xi}_2}+\Gamma_{12}\e^{\bar{\xi}_1+\xi_2}+\Gamma_{22}\e^{\xi_2+\bar{\xi}_2} \nonumber\\
&&~~~~~~~~~+\e^{\xi_1+2\bar{\xi}_1+\xi_2+\phi_1}+\e^{2\xi_1+\bar{\xi}_1+2\bar{\xi}_2+\phi_2}+\e^{\xi_1+\xi_2+2\bar{\xi}_2+\phi_3}+\e^{2\xi_2+\bar{\xi}_1+\bar{\xi}_2+\phi_4}\nonumber\\
&&~~~~~~~~~+\e^{\xi_1+\bar{\xi}_1+\xi_2+\bar{\xi}_2+\phi_5}+\e^{2\xi_1+2\bar{\xi}_1+\xi_2+\bar{\xi}_2+\phi_{11}}+\e^{\xi_1+2\bar{\xi}_1+2\xi_2+\bar{\xi}_2+\phi_{12}}+\e^{2\xi_1+\bar{\xi}_1+\xi_2+2\bar{\xi}_2+\phi_{13}}\nonumber\\
&&~~~~~~~~~+\e^{\xi_1+\bar{\xi}_1+2\xi_2+2\bar{\xi}_2+\phi_{14}}\nonumber
\eea
where the constants are obtained as
\bea
&&\e^{\phi_{1}}=\frac{-\varrho_{12}^2\Gamma_{11}\Gamma_{12}}{\kappa_{11}\kappa_{21}},~\e^{\phi_{2}}=\frac{-\bar{\varrho}_{12}^2\Gamma_{11}\Gamma_{21}}{\kappa_{11}\kappa_{12}},\e^{\phi_{3}}=\frac{-\varrho_{12}^2\Gamma_{21}\Gamma_{22}}{\kappa_{12}\kappa_{22}},~\e^{\phi_{4}}=\frac{-\bar{\varrho}_{12}^2\Gamma_{12}\Gamma_{22}}{\kappa_{21}\kappa_{22}},\nonumber\\
&&\e^{\phi_5}=\frac{\Gamma_{11}\Gamma_{22}(\kappa_{12}\kappa_{21})^{1/2}\Lambda_1+\Gamma_{12}\Gamma_{21}(\kappa_{11}\kappa_{22})^{1/2}\Lambda_2}{\kappa_{11}\kappa_{12}\kappa_{21}\kappa_{22}},\nonumber\\
&&\Lambda_1=(2\kappa_{11}(\kappa_{21}\kappa_{12})^{\frac{1}{2}}-\kappa_{11}\kappa_{12}^{\frac{1}{2}}(2\bar{k}_1+k_2-\bar{k}_2)-\kappa_{11}\kappa_{21}^{\frac{1}{2}}(2k_1-k_2+\bar{k}_2)\nonumber\\
&&~~~~~~~+\kappa_{21}^{\frac{1}{2}}\kappa_{22}(-k_1+\bar{k}_1-2\bar{k}_2)+(k_1-\bar{k}_1-2k_2)\kappa_{12}+2(\kappa_{12}\kappa_{21})^{\frac{1}{2}}\kappa_{22}),\nonumber\\
&&\Lambda_2=(\kappa_{11}^{\frac{1}{2}}(k_1-2k_2-\bar{k}_2)\kappa_{21}+\kappa_{11}^{\frac{1}{2}}\kappa_{12}(\bar{k}_1-k_2-2\bar{k}_2)+2(\kappa_{11}\kappa_{22})^{\frac{1}{2}}\kappa_{21}\nonumber\\
&&~~~~~~~ -\kappa_{21}\kappa_{22}^{\frac{1}{2}}(k_1+2\bar{k}_1-\bar{k}_2)+2\kappa_{12}\kappa_{22}\kappa_{11}^{\frac{1}{2}}-(2k_1+\bar{k}_1-k_2)\kappa_{12}\kappa_{22}^{\frac{1}{2}}),\nonumber\\
&&\e^{\phi_{11}}=\frac{\varrho_{12}\bar{\varrho}_{12}\Gamma_{11}\psi}{\kappa_{11}^2\kappa_{21}\kappa_{12}},~\e^{\phi_{12}}=\frac{\varrho_{12}\bar{\varrho}_{12}\Gamma_{12}\psi}{\kappa_{11}\kappa_{21}^2\kappa_{22}},~\e^{\phi_{13}}=\frac{\varrho_{12}\bar{\varrho}_{12}\Gamma_{21}\psi}{\kappa_{11}\kappa_{12}^2\kappa_{22}},~\e^{\phi_{14}}=\frac{\varrho_{12}\bar{\varrho}_{12}\Gamma_{22}\psi}{\kappa_{21}\kappa_{12}\kappa_{22}^2},\nonumber\\
&&\psi=\bigg(k_2\bar{k}_2\Gamma_{11}\Gamma_{22}+k_1(-\bar{k}_2\Gamma_{21}\Gamma_{12}+\bar{k}_1\Gamma_{11}\Gamma_{22}-k_2\nu_1\nu_2)-\bar{k}_1(k_2\Gamma_{21}\Gamma_{12}+\bar{k}_2\nu_1\nu_2)\bigg),\nonumber\eea

{\bf \section*{Conflicts of interest}}
The authors declare that they have no conflict of interest.


\begin{thebibliography}{}
%
%
\bibitem{a}
Suchkov SV,  Sukhorukov AA, Huang J, Dmitriev SV,  Lee C, Kivshar YS, Nonlinear switching and solitons in $\cal{PT}$-symmetric photonic systems, \emph{Laser Photonics Rev.},  {\bf 10} 177 (2016).

\bibitem{b}
Konotop VV, Yang J, Zezyulin DA, Nonlinear waves in $\cal{PT}$-symmetric systems, \emph{Rev. of Mod. Phys.},  {\bf 88},  035002 (2016).

\bibitem{3a}
Ablowitz MJ,  Musslimani ZH, Integrable nonlocal nonlinear Schr\"{o}dinger equation,  \emph{Phys. Rev. Lett.}, {\bf 110}, 064105 (2013). 

\bibitem{3b}
Ablowitz MJ,  Musslimani ZH,  Inverse scattering transform for the integrable nonlocal nonlinear Schr\"{o}dinger equation, \emph{Nonlinearity}, {\bf 29}, 915 (2016).


\bibitem{4a}
Gerdjikov VS, Saxena A, Complete integrability of nonlocal nonlinear Schr\"{o}dinger equation, \emph{J Math Phys}   {\bf 58}, 013502 (2017). 

\bibitem{1}
Bender CM, Boettcher S, Real Spectra in Non-Hermitian Hamiltonians Having $\cal{PT}$-Symmetry, \emph{Phys. Rev. Lett.}, {\bf 80}, 5243 (1998).

\bibitem{3}
Sarma AK,  Miri M A , Musslimani ZH, Christodoulides DN, Continuous and discrete Schr\"{o}dinger systems with parity-time-symmetric nonlinearities, \emph{Phys. Rev. E}, {\bf 89}, 052918 (2014). 

\bibitem{4b}
Gadzhimuradov  TA, Agalarov AM, Towards a gauge-equivalent magnetic structure of the nonlocal nonlinear Schr\"{o}dinger equation,   \emph{Phys. Rev. A}, {\bf 93}, 062124 (2016). 
\bibitem{4c}
Lakshmanan M, Continuum spin system as an exactly solvable dynamical system, \emph {Phys. Lett. A}, {\bf 61}, 53 (1977).
\bibitem{5a}
Khare A, Saxena A, Periodic and hyperbolic soliton solutions of a number of nonlocal nonlinear equations, \emph{J. Math. Phys.} {\bf 56}, 032104 (2015). 

\bibitem{5b}
Li M, Xu T, Dark and antidark soliton interactions in the nonlocal nonlinear Schr\"{o}dinger equation with the self-induced parity-time-symmetric potential, \emph{Phys. Rev. E},  {\bf 91}, 033202 (2015). 

\bibitem{5c}
Huang X, Ling L, Soliton solutions for the nonlocal nonlinear Schr\"{o}dinger equation,  \emph{Eur. Phys. J. Plus}, {\bf 131}, 148 (2016).

\bibitem{5d}
Wen X Y, Yan Z, Yang Y, Dynamics of higher-order rational solitons for the nonlocal nonlinear Schr\"{o}dinger equation with the self-induced parity-time-symmetric potential,  \emph{Chaos},  {\bf 26}, 063123 (2016). 

\bibitem{5e}
Stalin S, Senthilvelan M, Lakshmanan M, Nonstandard bilinearization of $\cal{PT}$-invariant nonlocal nonlinear Schr\"{o}dinger equation: Bright soliton solutions, \emph{ Phys. Lett. A},  {\bf 381},  2380 (2017). 

\bibitem{51}
Ablowitz MJ, Musslimani ZH, Integrable discrete $\cal{PT}$-symmetric model, \emph {Phys. Rev. E},  {\bf 90}, 032912 (2014). 

\bibitem{52}
Fokas  AS, Integrable multidimensional versions of the nonlocal nonlinear Schr\"{o}dinger equation, \emph{Nonlinearity}, {\bf 29}, 319 (2016).   

\bibitem{532}
Lou SY, Huang F, Alice-Bob Physics: Coherent Solutions of Nonlocal KdV Systems, \emph{Sci. Rep.},  {\bf 7}, 869 (2017).

\bibitem{533}
Rao J, Cheng Y, He J, Rational and Semirational Solutions of the Nonlocal Davey–Stewartson Equations,\emph{Stud. Appl. Math},  {\bf 139}, 568 (2017).

\bibitem{534}
Liu Y, Mihalache D,  He J, Families of rational solutions of the y-nonlocal Davey–Stewartson II equation, \emph{Nonlinear Dyn.}, {\bf 90}, 2445 (2017).

\bibitem{54}
Xu Z X, Chow K W, Breathers and rogue waves for a third order nonlocal partial differential equation by a bilinear transformation, \emph{Appl. Math. Lett.},  {\bf 56}, 72 (2016). 

\bibitem{55}
Li M, Xu T, Meng D, Reverse Space-Time Nonlocal Sasa-Satsuma Equation and Its Solutions, \emph{J. Phys. Soc. Jpn},  {\bf 85}, 124001 (2016). 
\bibitem{54}
 Ma LY,  Zhu ZN,  $N$-soliton solution for an integrable nonlocal discrete focusing nonlinear Schrödinger equation, \emph{Appl. Math. Lett.}, {\bf 59}, 115 (2016).
\bibitem{55}
Ji JL, Zhu ZN, On a nonlocal modified Korteweg-de Vries equation: Integrability, Darboux transformation and soliton solutions, \emph{ Commun. Nonlinear Sci. Numer. Simul.},  {\bf 42}, 699 (2017).
\bibitem{56}
 Ma LY, Zhu ZN, Nonlocal nonlinear Schr\"{o}dinger equation and its discrete version: Soliton solutions and gauge equivalence, \emph{J. Math. Phys.},  {\bf 57}, 083507 (2016).
\bibitem{57}
Zhang HQ, Zhang MY, Hu R, Darboux transformation and soliton solutions in the parity-time-symmetric nonlocal vector nonlinear Schr\"{o}dinger equation, \emph{Appl. Math. Lett.},  {\bf 76}, 170 (2018). 
\bibitem{58}
Wen Z, Yan Z, Solitons and their stability in the nonlocal nonlinear Schr\"{o}dinger equation with $\cal{PT}$-symmetric potentials, \emph{Chaos},  {\bf  27}, 053105 (2017).

\bibitem{60}
Zhang G, Yan Z, Multi-rational and semi-rational solitons and interactions for the nonlocal coupled nonlinear Schr\"{o}dinger equations, \emph{Euro. Phys. Lett.}, {\bf 118}, 60004 ( 2017). 

\bibitem{61}
Chen K, Zhang D J,  Solutions of the nonlocal nonlinear Schr\"{o}dinger hierarchy via reduction, \emph{Appl. Math. Lett.}, {\bf 75}, 82 (2018).
\bibitem{62}
Chen K, Deng X, Lou S, Zhan D J, Solutions of Nonlocal Equations Reduced from the AKNS Hierarchy, \emph{Stud. Appl. Math.}, {\bf 00}, 1 (2018).

\bibitem{63}
Liu W, Li X , General soliton solutions to a $(2+1)$-dimensional nonlocal nonlinear Schr\"{o}dinger equation with zero and nonzero boundary conditions, \emph{Nonlinear Dyn.}, (2018). https://doi.org/10.1007/s11071-018-4221-2. 

\bibitem{64}
 Tang XY,  Liang Z F, A general nonlocal nonlinear Schr\"{o}dinger equation with shifted parity, charge-conjugate and delayed time reversal, \emph{Nonlinear Dyn.}, {\bf 92}, 815 (2018). 

\bibitem{65}
 Zhang Y, Liu Y, Tang X , A general integrable three-component coupled nonlocal nonlinear Schrödinger equation, \emph{Nonlinear Dyn.}, {\bf 89 }, 2729 (2017). 

\bibitem{66}
Sun B, General soliton solutions to a nonlocal long-wave–short-wave resonance interaction equation with nonzero boundary condition, \emph{Nonlinear Dyn.}, {\bf 92}, 1369 (2018).

\bibitem{67}
Ma L Y, Zhao H Q, Gu H, Integrability and gauge equivalence of the reverse space–time nonlocal Sasa–Satsuma equation, \emph{Nonlinear Dyn.}, {\bf 91}, 1909 (2018).
    
\bibitem{6a}
Radhakrishnan R, Lakshmanan M, Hietarinta J, Inelastic collision and switching of coupled bright solitons in optical fibers, \emph { Phys. Rev. E},   {\bf 56}, 2213 (1997).
\bibitem{6b}
Kanna T, Lakshmanan M, Exact soliton solutions, shape changing collisions, and partially coherent solitons in coupled nonlinear Schr\"{o}dinger equations,  {\it Phys. Rev. Lett.}, {\bf 86},  5043 (2001). 

\bibitem{6c}
Radhakrishnan R, Lakshmanan M, Bright and dark soliton solutions to coupled nonlinear Schr\"{o}dinger equations,  {\it J. Phys. A. Math. Theor.},  {\bf 28} 2683 (1995).
\bibitem{6d}
Sheppard AP, Kivshar YS, Polarized dark solitons in isotropic Kerr media,  {\it Phys. Rev. E},  {\bf 55}, 4773 (1997). 

\bibitem{7a}
Kanna T, Lakshmanan M, Exact soliton solutions of coupled nonlinear Schr\"{o}dinger equations: Shape-changing collisions, logic gates, and partially coherent solitons,  {\it Phys. Rev. E}, {\bf 67}, 046617 ( 2003);
\bibitem{7a1}  
Vijayajayanthi M, Kanna T, Lakshmanan M, Bright-dark solitons and their collisions in mixed $N$-coupled nonlinear Schr\"{o}dinger equations, \emph {Phys. Rev. A}, {\bf 77}, 013820 (2008). 
\bibitem{7a2}
Kanna T, Lakshmanan M, Dinda PT, Akhmediev N, Soliton collisions with shape change by intensity redistribution in mixed coupled nonlinear Schr\"{o}dinger equations, \emph {Phys. Rev. E}, {\bf 73}, 026604 (2006). 
\bibitem {7b}
Ohta Y, Wang DS, Yang J, General N‐Dark–Dark Solitons in the Coupled Nonlinear Schrödinger Equations,  \emph {Stud. Appl. Math}, {\bf 127}, 345 (2011).

\bibitem{8a}
Sinha D, Ghosh PK, Integrable nonlocal vector nonlinear Schr\"{o}dinger equation with self-induced parity-time-symmetric potential, \emph{Phys. Lett. A},  {\bf 381}, 124 (2017). 

\bibitem{6}
Gilson C, Hietarinta J, Nimmo J, Ohta Y, Sasa-Satsuma higher-order nonlinear Schr\"{o}dinger equation and its bilinearization and multisoliton solutions,  \emph{Phys. Rev. E},  {\bf 68}, 016614 (2003). 

\bibitem{7}
Kanna T, Vijayajayanthi M, Lakshmanan M, Coherently coupled bright optical solitons and their collisions, \emph{J. Phys. A: Math. Theor.}, {\bf 43}, 434018 (2010). 

\bibitem{8}
Hirota  R, The Direct Method in soliton Theory, Cambridge University Press,  Cambridge (2004).


\bibitem{7f}
Li S, Biondini G, Schiebold C, On the degenerate soliton solutions of the focusing nonlinear Schrödinger equation,  \emph{J. Math. Phys.},  {\bf 58}, 033507 (2017).

\bibitem{7g}
Cen J , Correa F, Fring A, Degenerate multi-solitons in the sine-Gordon equation, \emph{J. Phys. A: Math. Theor.}, {\bf 50},  435201 (2017). 


\end{thebibliography}
\end{document}